\documentclass[a4paper,11pt]{article}
\pdfoutput=1 % if your are submitting a pdflatex (i.e. if you have
\usepackage{jheppub} % for details on the use of the package, please
\usepackage[T1]{fontenc} % if needed
\usepackage{arydshln}

\usepackage{rotating}
\usepackage{hhline}
\usepackage{multirow}

\usepackage{feynmp}
\allowdisplaybreaks

\def\beq{\begin{equation}}
\def\eeq#1{\label{#1}\end{equation}}
\def\eeqn{\end{equation}}

\def\beqa{\begin{eqnarray}}
\def\eeqa#1{\label{#1}\end{eqnarray}}
\def\eeqan{\end{eqnarray}}
\def\CR{\nonumber \\ }

\def\bseq{\begin{subequations}}
\def\eseq#1{\label{#1}\end{subequations}}
\def\eseqn{\end{subequations}}

\def\Dlr{\overleftrightarrow{D}}
\def\VCKM{V_{\text{CKM}}}
\def\L{{\cal L}}
\def\O{{\cal O}}

\def\SM{\text{SM}}

\def\cw{c_\theta}
\def\sw{s_\theta}
\def\cwb{\bar{c}_\theta}
\def\swb{\bar{s}_\theta}
\def\cwh{\hat{c}_\theta}
\def\swh{\hat{s}_\theta}
\def\Dgzb{\Delta\bar g_1^Z}

\def\Dkapzb{\Delta\bar\kappa_Z}
\def\Dkapab{\Delta\bar\kappa_\gamma}

\def\lamab{\bar\lambda_\gamma}
\def\lamgb{\bar\lambda_g}
\def\DkapFb{\Delta\bar\kappa_F}
\def\DkapVb{\Delta\bar\kappa_V}
\def\De{\Delta\epsilon}
\def\dgL{\delta g_L}
\def\dgR{\delta g_R}
\def\dgz{\delta g_{1z}}
\def\dkapa{\delta\kappa_\gamma}

\def\muew{\mu_{\text{EW}}}
\def\ddlm{\frac{d}{d\ln\mu}}
\def\dgLd{\delta\dot g_L}
\def\dgRd{\delta\dot g_R}
\def\Rl{R_\ell}
\def\Rb{R_b}
\def\Gamhad{\Gamma_\text{had}}

\def\dbNP{\bar\delta^{\text{NP}}}
\def\Obs{\hat\O}

\preprint{MCTP-15-30}

\title{\boldmath Renormalization group evolution of the universal theories EFT}

\author{James~D.~Wells}
\author{and Zhengkang~Zhang}
\affiliation{Michigan Center for Theoretical Physics, Department of Physics, University of Michigan,\\Ann Arbor, MI 48109, U.S.A.}

\emailAdd{jwells@umich.edu}
\emailAdd{zzkevin@umich.edu}

\abstract{The conventional oblique parameters analyses of precision electroweak data can be consistently cast in the modern framework of the Standard Model effective field theory (SMEFT) when restrictions are imposed on the SMEFT parameter space so that it describes universal theories. However, the usefulness of such analyses is challenged by the fact that universal theories at the scale of new physics, where they are matched onto the SMEFT, can flow to nonuniversal theories with renormalization group (RG) evolution down to the electroweak scale, where precision observables are measured. The departure from universal theories at the electroweak scale is not arbitrary, but dictated by the universal parameters at the matching scale. But to define oblique parameters, and more generally universal parameters at the electroweak scale that directly map onto observables, additional prescriptions are needed for the treatment of RG-induced nonuniversal effects. 
We perform a RG analysis of the SMEFT description of universal theories, and discuss the impact of RG on simplified, universal-theories-motivated approaches to fitting precision electroweak and Higgs data.}

\begin{document} 
\maketitle
\flushbottom

\section{Introduction}

The quest for new physics beyond the Standard Model (BSM) has been, and will continue to be proceeding through both direct and indirect searches for their effects. While direct searches for BSM signatures have to be carried out with particular models (often simplified ones) in mind, indirect searches through precision measurements of Standard Model (SM) processes often admit more general approaches that are model-independent to some extent. A classic example is the oblique parameters formalism~\cite{Kennedy:1988sn}, the widely-adopted version of which was proposed by Peskin and Takeuchi~\cite{Peskin:1991sw}, and further developed by others~\cite{Maksymyk:1993zm,Barbieri:2004qk}. Here, just a few parameters, most notably $S$ and $T$ (or their rescaled versions $\hat S$ and $\hat T$), capture the new physics modifications of the vector boson self-energies, which are assumed to be the dominant BSM effects (hence the name ``oblique''). Modern studies in this direction are migrating to the Standard Model effective field theory (SMEFT) approach; see e.g.~\cite{Degrande:2012wf,Willenbrock:2014bja,Falkowski:2015fla} for recent reviews. In this case, the SM Lagrangian, supplemented by the complete set of dimension-6 operators built from the SM field content, provides a most general and consistent framework for calculating the leading BSM effects on precision observables, assuming there are no new light states and the new physics scale $\Lambda$ is much higher than the electroweak scale $\muew$.

Reconciliation of the oblique parameters formalism and the more general SMEFT is based on the realization that the former is generally speaking only applicable to universal theories, a restricted class of BSM theories whose SMEFT representation can be cast in a form that involves bosonic operators only~\cite{Wells:2015uba} (see also~\cite{Trott:2014dma} for an earlier study with similar motivations). By bosonic operators, we mean dimension-6 operators built from the SM bosons. There are 16 of them one can possibly write down that are independent and CP-even, as we have shown in~\cite{Wells:2015uba}, so the effective theory of universal theories has a 16-dimensional parameter space, independent of the SMEFT basis choice. In turn, they can be mapped onto 16 independent phenomenological parameters, called {\it``universal parameters''} in~\cite{Wells:2015uba}, 5 of which coincide with the familiar oblique parameters. At leading order (LO) in $\frac{v^2}{\Lambda^2}$, they lead to a universal pattern of deviations from the SM. In the recently-proposed Higgs basis framework~\cite{HiggsBasis}, this pattern is encoded in a set of relations among the otherwise independent effective couplings.

\begin{figure}[tbp]
\centering % \begin{center}/\end{center} takes some additional vertical space
%\includegraphics[width=.45\textwidth,trim=0 380 0 200,clip]{img1.pdf}
%\hfill
%\includegraphics[width=.45\textwidth,origin=c,angle=180]{img2.pdf}
% "\includegraphics" is very powerful; the graphicx package is already loaded
%\begin{subfigure}
\begin{fmffile}{nudiag}
\begin{fmfgraph*}(100,100)
\fmfleft{w}
\fmfright{f1,o,f2}
\fmfblob{.15w}{b}
\fmf{photon,label=$W^\pm$,l.side=left}{w,wff}
\fmf{fermion}{f1,vff1,wff,vff2,f2}
\fmffreeze
%\fmf{photon,right=0.3,label=$Z/\gamma$}{vff1,b,vff2}
\fmf{photon,right=0.3}{vff1,b}
\fmf{photon,right=0.3,label=$Z/\gamma$}{b,vff2}
\fmf{phantom}{b,o}
\end{fmfgraph*}
\hspace{0.5in}
\begin{fmfgraph*}(100,100)
\fmfleft{o1,z,o2}
\fmfright{f1,f2}
\fmfblob{.15w}{b}
\fmf{photon,label=$Z$,l.side=left,tension=2}{z,b}
\fmf{photon,tension=2}{b,vff}
%\fmf{photon,label={\footnotesize $Z/\gamma$},l.side=left}{b,vff}
\fmf{fermion,label=$b_L$,l.side=left}{f1,gff1}
\fmf{fermion,label=$t_L$,l.side=left}{gff1,vff}
\fmf{fermion,label=$t_L$,l.side=left}{vff,gff2}
\fmf{fermion,label=$b_L$,l.side=left}{gff2,f2}
%\fmf{fermion}{f1,vff1,wff,vff2,f2}
\fmffreeze
\fmf{dashes,right=0.5,label=$\phi^\pm$}{gff1,gff2}
\fmf{phantom}{b,o1}
\fmf{phantom}{b,o2}
\end{fmfgraph*}
\hspace{0.5in}
\begin{fmfgraph*}(100,100)
\fmfleft{h}
\fmfright{f1,o,f2}
\fmfblob{.15w}{b}
\fmf{dashes,label=$h$,l.side=left}{h,hff}
\fmf{fermion}{f1,vff1,hff,vff2,f2}
\fmffreeze
\fmf{photon,right=0.3}{vff1,b,vff2}
\fmf{phantom}{b,o}
\end{fmfgraph*}
\end{fmffile}
\caption{\label{fig:diag} Examples showing how nonuniversal effects can be generated by universal oblique corrections. {\bf Left:} effective $Wqq'$ and $W\ell\nu$ couplings are renormalized differently, due to the different couplings of quarks and leptons to neutral gauge bosons. {\bf Middle:} the $Zb_L\bar b_L$ coupling is singled out among all the $Zf\bar f$ couplings probed by $Z$-pole measurements for relatively large running effects proportional to $y_t^2$, via loop corrections involving the charged Goldstone boson (or the longitudinal $W^\pm$ if one uses the unitary gauge). {\bf Right:} the Higgs boson couplings to the up- and down-type quarks and leptons are renormalized differently, due to different gauge interactions of the fermions. In each example, the interactions generated for the SM fermions are not in the form of the SM currents, and thus the corresponding operators cannot be eliminated in favor of bosonic operators. These examples, as well as many others, can be more rigorously formulated in terms of $SU(2)_L\times U(1)_Y$ invariant operators, but we prefer to give a more intuitive illustration at this stage. The arguments here will be made concrete in sections~\ref{sec:ew} and~\ref{sec:yukawa}.}
\end{figure}
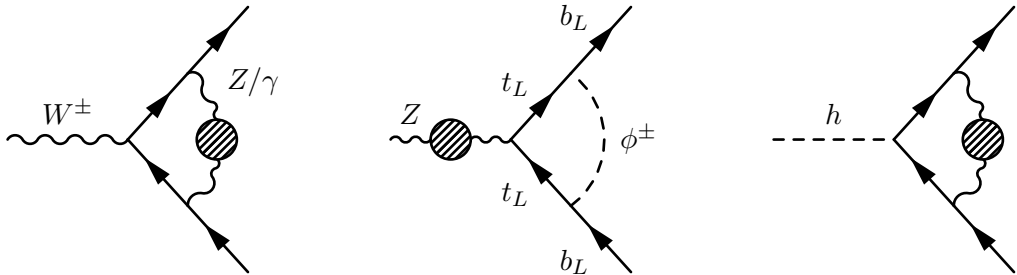

Beyond LO, however, complications can arise. In particular, the 16-dimensional parameter space of universal theories, being a subspace of the full SMEFT parameter space, is not guaranteed to be closed under renormalization group (RG) evolution. In fact, it is intuitively clear that nonuniversal effects can indeed be generated by RG, because even if one starts with a bosonic basis (consisting of 16 independent bosonic operators)~\cite{Wells:2015uba}, fermionic operators, i.e.\ operators containing SM fermions, can be generated that are not organized into the SM currents and hence cannot be eliminated in favor of bosonic operators. Three examples involving oblique corrections are illustrated in figure~\ref{fig:diag}. This qualitative argument can be made concrete by a detailed RG analysis of universal theories, which we perform in this paper,\footnote{It should be noted that in the SMEFT framework, observables at the electroweak scale are calculated as a double series expansion, in powers of both $\frac{E^2}{\Lambda^2}\sim\frac{v^2}{\Lambda^2}$ and the loop factor $\frac{1}{16\pi^2}$. Terms of order $(\frac{v^2}{\Lambda^2})^0(\frac{1}{16\pi^2})^n$ can be taken into account by incorporating higher-order SM calculations independently of new physics contributions~\cite{Wells:2014pga}. The LO new physics effects, like those discussed in~\cite{Wells:2015uba}, are of order $(\frac{v^2}{\Lambda^2})^1(\frac{1}{16\pi^2})^0$. The RG effects analyzed in the present paper correspond to order $(\frac{v^2}{\Lambda^2})^1(\frac{1}{16\pi^2})^1$ terms in the double expansion that are enhanced by $\ln\frac{\Lambda}{\muew}$. Terms of order $(\frac{v^2}{\Lambda^2})^2(\frac{1}{16\pi^2})^0$ may also have an impact, but the effective Lagrangian has to be extended beyond the dimension-6 level to account for them. The latter~\cite{Lehman:2015via,Lehman:2015coa} is beyond the scope of the present work. See also~\cite{Buchalla:2014eca,Berthier:2015oma,Berthier:2015gja} for related discussions.} aided by the recently-calculated full anomalous dimension matrix for the dimension-6 operators~\cite{Jenkins:2013zja,Jenkins:2013wua,Alonso:2013hga} (see also~\cite{Alonso:2014zka}).

As a consequence of the RG-induced nonuniversal effects, an effective theory that is universal at the new physics scale $\Lambda$ can become nonuniversal at the electroweak scale $\muew$. This means that, without introducing further prescriptions, the universal parameters $\hat S$, $\hat T$, etc.\ are not unambiguously defined beyond LO at the electroweak scale. However, the usefulness of these parameters is not plagued, since after all, their values at the high scale $\Lambda$ are what we really need to know to infer the shape of BSM physics. The latter are well-defined in universal theories, and the 16 of them are sufficient to describe phenomenology also at $\muew$, despite the theory becoming nonuniversal after RG evolution. Departures from universal BSM effects are not arbitrary as in generic nonuniversal theories, but can be calculabled in terms of these parameters. 
%Unlike in generic nonuniversal theories, departures from universal BSM effects follow a special pattern controlled by the 16 universal parameters at $\Lambda$.
%Unlike generic nonuniversal theories, the theories that universal theories flow to are in a sense ``universally nonuniversal,'' where departures from universal BSM effects are determined by the 16 universal parameters at $\Lambda$.
%Unlike generic nonuniversal theories, the nonuniversal theories at the electroweak scale that universal theories flow to still exhibit a universal pattern -- they are still controlled by the 16 universal parameters at $\Lambda$, and are in this sense ``universally nonuniversal.''

An important motivation for the recent trend to push the SMEFT analyses beyond LO~\cite{Jenkins:2013zja,Jenkins:2013wua,Alonso:2013hga,Alonso:2014zka,Masso:2012eq,Corbett:2012ja,Grojean:2013kd,Elias-Miro:2013gya,Falkowski:2013dza,Contino:2013kra,Jenkins:2013fya,Mebane:2013zga,Elias-Miro:2013mua,Pomarol:2013zra,Chen:2013kfa,Elias-Miro:2013eta,Belusca-Maito:2014dpa,Englert:2014cva,Henning:2014wua,Berthier:2015oma,Drozd:2015kva,Hartmann:2015oia,Ghezzi:2015vva,Chiang:2015ura,Huo:2015exa,deBlas:2015aea,Hartmann:2015aia,Berthier:2015gja,Huo:2015nka,David:2015waa,Gauld:2015lmb,Drozd:2015rsp} (see also~\cite{Buchalla:2012qq,Buchalla:2013rka,Brivio:2013pma,Buchalla:2013eza,Gavela:2014uta,Gonzalez-Alonso:2014eva,Buchalla:2014eca,Bordone:2015nqa,Buchalla:2015qju}) is the observation that for some very well-measured observables, it is possible to derive additional constraints on the effective operators contributing at higher loop order, which are otherwise less constrained.\footnote{Note, however, that in some of these references, bounds on the oblique parameters are used to constrain the SMEFT parameter space possibly beyond the universal theories subspace, which can lead to inconsistencies as argued in~\cite{Wells:2015uba} (see also~\cite{Trott:2014dma}). The results should therefore be interpreted with caution.} In the full SMEFT, this can be done at the leading logarithmic (LL) level by first constraining the Wilson coefficients at $\muew$ via LO expressions of the observables at the electroweak scale, and then RG-evolving these constraints to $\Lambda$. The same is not true for the universal parameters $\hat S$, $\hat T$, etc. As we will see, with additional prescriptions, it is possible to define these parameters at $\muew$, but they do not capture all the LL corrections to all observables no matter what prescriptions are adopted. This implies, in particular, that the conventional oblique parameters analysis incorporating only LO effects of the oblique parameters is not a priori justified at the LL level, where additional parameters that should have been included in the fit may have a numerical impact. 
%This implies, in particular, that the conventional oblique parameters analysis of precision electroweak data is not a priori justified beyond LO accuracy, if no additional parameters are included in the fit to account for the remaining LL corrections. 
Also, a simplified global fit to Higgs data where a single rescaling parameter $\DkapFb$ is assumed for all the $hf\bar f$ couplings may not be appropriate, since it may not even accurately capture the phenomenology of universal theories.

The paper is organized as follows. We begin in section~\ref{sec:ut} with a brief review of the universal theories EFT at LO, and a general discussion of RG-induced nonuniversal effects. Sections~\ref{sec:ew} and~\ref{sec:yukawa} are devoted to detailed RG analyses of universal theories in the electroweak and Yukawa sectors, respectively. We conclude in section~\ref{sec:conclusions}, and collect our notation and some useful formulas in the appendices.

%%%
\section{Universal theories at LO and beyond}
\label{sec:ut}

%%%
\subsection{The universal theories EFT at LO}
\label{sec:lo}

In this subsection, we briefly review the results in~\cite{Wells:2015uba}. The SMEFT description of universal theories at LO can be formulated in three equivalent ways, in terms of effective operators, universal parameters, or Higgs basis couplings.

\begin{table}[tbp]
\centering
\begin{tabular}{|c@{$\,\equiv\,$}l|l|}
\hline
\multicolumn{2}{|c|}{Definition} & \multicolumn{1}{c|}{Warsaw basis operator combination} \\
\hline
$Q_{HJW}$ & $\frac{ig}{2}(H^\dagger\sigma^a\Dlr_\mu H)J_W^{a\mu}$ & $\frac{1}{4}g^2 \bigl([Q_{Hq}^{(3)}]_{ii} +[Q_{Hl}^{(3)}]_{ii}\bigr)$ \\
\hline
$Q_{HJB}$ & $\frac{ig'}{2}(H^\dagger\Dlr_\mu H)J_B^\mu$ & $\frac{1}{2}g'^2 \bigl(Y_q[Q_{Hq}^{(1)}]_{ii} +Y_l[Q_{Hl}^{(1)}]_{ii}$ \\
\multicolumn{2}{|c|}{} & $\quad +Y_u[Q_{Hu}]_{ii} +Y_d[Q_{Hd}]_{ii} +Y_e[Q_{He}]_{ii}\bigr)$ \\
\hline
$Q_{2JW}$ & $J_{W\mu}^aJ_W^{a\mu}$ & $g^2\bigl(\frac{1}{4}[Q_{qq}^{(3)}]_{iijj} -\frac{1}{4}[Q_{ll}]_{iijj} +\frac{1}{2}[Q_{ll}]_{ijji} +\frac{1}{2}[Q_{lq}^{(3)}]_{iijj} \bigr)$ \\
\hline
$Q_{2JB}$ & $J_{B\mu}J_B^\mu$ & $g'^2 \bigl( Y_q^2[Q_{qq}^{(1)}]_{iijj} +Y_l^2[Q_{ll}]_{iijj} +2Y_qY_l[Q_{lq}^{(1)}]_{iijj}$ \\
\multicolumn{2}{|c|}{} & $\quad +Y_u^2[Q_{uu}]_{iijj} +Y_d^2[Q_{dd}]_{iijj} +Y_e^2[Q_{ee}]_{iijj}$ \\
\multicolumn{2}{|c|}{} & $\quad +2Y_qY_u[Q_{qu}^{(1)}]_{iijj} +2Y_qY_d[Q_{qd}^{(1)}]_{iijj} +2Y_qY_e[Q_{qe}]_{iijj}$ \\
\multicolumn{2}{|c|}{} & $\quad +2Y_lY_u[Q_{lu}]_{iijj} +2Y_lY_d[Q_{ld}]_{iijj} +2Y_lY_e[Q_{le}]_{iijj}$ \\
\multicolumn{2}{|c|}{} & $\quad +2Y_uY_d[Q_{ud}^{(1)}]_{iijj} +2Y_uY_e[Q_{eu}]_{iijj} +2Y_dY_e[Q_{ed}]_{iijj} \bigr)$ \\
\hline
$Q_{2JG}$ & $J_{G\mu}^AJ_G^{A\mu}$ & $g_s^2 \bigl( -\frac{1}{6}[Q_{qq}^{(1)}]_{iijj} +\frac{1}{4}[Q_{qq}^{(1)}]_{ijji} +\frac{1}{4}[Q_{qq}^{(3)}]_{ijji}$ \\
\multicolumn{2}{|c|}{} & $\quad -\frac{1}{6}[Q_{uu}]_{iijj} +\frac{1}{2}[Q_{uu}]_{ijji} -\frac{1}{6}[Q_{dd}]_{iijj} +\frac{1}{2}[Q_{dd}]_{ijji}$ \\
\multicolumn{2}{|c|}{} & $\quad +2[Q_{qu}^{(8)}]_{iijj} +2[Q_{qd}^{(8)}]_{iijj} +2[Q_{ud}^{(8)}]_{iijj} \bigr)$ \\
\hline
$Q_y$ & $|H|^2 (H_\alpha J_y^\alpha+\text{h.c.})$ & $[y_u]_{ij} [Q_{uH}]_{ij} + [\VCKM y_d]_{ij} [Q_{dH}]_{ij} + [y_e]_{ij} [Q_{eH}]_{ij} +\text{h.c.}$ \\
\hline
$Q_{2y}$ & $J_{y\alpha}^\dagger J_y^\alpha$ & $-[y_u]_{il}[y_u^\dagger]_{kj}\bigl(\frac{1}{6}[Q_{qu}^{(1)}]_{ijkl}+[Q_{qu}^{(8)}]_{ijkl}\bigr) -\frac{1}{2}[y_e]_{il}[y_e^\dagger]_{kj}[Q_{le}]_{ijkl}$ \\
\multicolumn{2}{|c|}{} & $\quad -[\VCKM y_d]_{il}[y_d^\dagger\VCKM^\dagger]_{kj}\bigl(\frac{1}{6}[Q_{qd}^{(1)}]_{ijkl}+[Q_{qd}^{(8)}]_{ijkl}\bigr)$ \\
\multicolumn{2}{|c|}{} & $\quad +\bigl( [y_u]_{ij}[\VCKM y_d]_{kl}[Q_{quqd}^{(1)}]_{ijkl} -[y_e]_{ij}[y_u]_{kl}[Q_{lequ}^{(1)}]_{ijkl}$ \\
\multicolumn{2}{|c|}{} & $\qquad +[y_e]_{ij}[y_d^\dagger\VCKM^\dagger]_{kl}[Q_{ledq}]_{ijkl} +\text{h.c.} \bigr)$ \\
\hline
\end{tabular}
\caption{\label{tab:Qf} Warsaw basis operator combinations that appear in $\L_\text{universal}$ in \eqref{Lu}. In these expressions, repeated generation indices $i,j,k,l$ are summed over, $H^\dagger\sigma^a\protect\Dlr_\mu H = H^\dagger\sigma^a (D_\mu H) - (D_\mu H)^\dagger\sigma^a H$, $H^\dagger\protect\Dlr_\mu H = H^\dagger (D_\mu H) - (D_\mu H)^\dagger H$. The Yukawa matrices $y_u$, $y_d$, $y_e$ should not be confused with the hypercharges $Y_f$. The SM vector and scalar currents $J_{G\mu}^A$, $J_{W\mu}^a$, $J_{B\mu}$, $J_y^\alpha$ are defined in \eqref{Jdef}. See appendix~\ref{app:notation} for definitions of the operators appearing in this table.}
\end{table}

As mentioned in the introduction, the effective Lagrangian of universal theories consists of $\L_\SM$ plus 16 independent CP-even bosonic operators. In the Warsaw basis~\cite{Grzadkowski:2010es}, only 9 of them are kept, while the remaining bosonic operators are eliminated by field redefinitions, or equivalently, by applying the SM equations of motion, in favor of combinations of fermionic operators. Despite the appearance of a proliferation of fermionic operators, the number of independent parameters (Wilson coefficients) is still 16. To be specific, using the notation of~\cite{Grzadkowski:2010es} for the Warsaw basis operators $Q_i$, collected in appendix~\ref{app:notation}, we have
\beqa
%\L_{\text{universal}} &=& \L_{\text{SM}} + \frac{1}{v^2} (C_{HW}Q_{HW} + C_{HB}Q_{HB} + C_{HG}Q_{HG} + C_{HWB}Q_{HWB} \CR
%&& + C_{W}Q_{W} +C_{G}Q_{G} + C_{HD}Q_{HD} + C_{H\square}Q_{H\square} + C_HQ_H \CR
%&& +C_{HJW}Q_{HJW} +C_{HJB}Q_{HJB} + C_{2JW}Q_{2JW} + C_{2JB}Q_{2JB} + C_{2JG}Q_{2JG} \CR
%&& + C_yQ_y + C_{2y}Q_{2y}),
\L_{\text{universal}} &=& \L_{\text{SM}} + \frac{1}{v^2} (C_{HW}Q_{HW} + C_{HB}Q_{HB} + C_{HG}Q_{HG} + C_{HWB}Q_{HWB} + C_{W}Q_{W} \CR
&& +C_{G}Q_{G} + C_{HD}Q_{HD} + C_{H\square}Q_{H\square} + C_HQ_H +C_{HJW}Q_{HJW} +C_{HJB}Q_{HJB} \CR
&& + C_{2JW}Q_{2JW} + C_{2JB}Q_{2JB} + C_{2JG}Q_{2JG} + C_yQ_y + C_{2y}Q_{2y}),
\eeqa{Lu}
where $C_i$ are $\O(\frac{v^2}{\Lambda^2})$ Wilson coefficients, and $Q_{HJW}$, $Q_{HJB}$, $Q_{2JW}$, $Q_{2JB}$, $Q_{2JG}$, $Q_{y}$, $Q_{2y}$ are combinations of fermionic operators listed in table~\ref{tab:Qf}. Note that the SM fermion fields appear in these combinations via the vector and scalar currents in the SM,
\bseq
\beqa
J_{G\mu}^A &\equiv& g_s\sum_{f\in\{q,u,d\}} \bar f \gamma_\mu T^A f, \\
J_{W\mu}^a &\equiv& g\sum_{f\in\{q,l\}} \bar f \gamma_\mu \frac{\sigma^a}{2} f, \\
J_{B\mu} &\equiv& g'\sum_{f\in\{q,l,u,d,e\}} Y_f \bar f \gamma_\mu f, \\
J_y^\alpha &\equiv& \bar u y_u^\dagger q_\beta \epsilon^{\beta\alpha} + \bar q^\alpha \VCKM y_d d + \bar l^\alpha y_e e.
\eeqan
\eseq{Jdef}
Our notation is such that
\beq
\L_\text{SM} \supset G^{A\mu}J_{G\mu}^A + W^{a\mu}J_{W\mu}^a + B^\mu J_{B\mu} - (H_\alpha J_y^\alpha+\text{h.c.}).
\eeqn
See appendix~\ref{app:notation} for more details. We will stick to the Warsaw basis for the calculations in this paper, in order to conveniently use the results in~\cite{Jenkins:2013zja,Jenkins:2013wua,Alonso:2013hga}. The forms of $\L_\text{universal}$ in other SMEFT bases, as well as the dictionaries for translating between the bases for universal theories, can be found in~\cite{Wells:2015uba}.

\begin{table}[tbp]
\centering
\begin{tabular}{|c|l|}
\hline
Universal & \multirow{2}{3.75in}{\centering Warsaw basis expression} \\
parameter & \\
\hline
$\hat S$ & $g^2\bigl(\frac{1}{gg'}C_{HWB} +\frac{1}{4}C_{HJW} +\frac{1}{4}C_{HJB} -\frac{1}{2}C_{2JW} -\frac{1}{2}C_{2JB}\bigr)$ \\
$\hat T$ & $-\frac{1}{2}C_{HD} +\frac{g'^2}{2}(C_{HJB}-C_{2JB})$ \\
$W$ & $-\frac{g^2}{2}C_{2JW}$ \\
$Y$ & $-\frac{g^2}{2}C_{2JB}$ \\
$Z$ & $-\frac{g^2}{2}C_{2JG}$ \\
\hline
$\Dgzb$ & $-\frac{g^2}{4\cw^2}(C_{HJW}-2C_{2JW})$ \\
$\Dkapab$ & $\frac{\cw}{\sw}C_{HWB}$ \\
$\lamab$ & $-\frac{3g}{2}C_W$ \\
$\lamgb$ & $-\frac{3g^2}{2g_s}C_G$ \\
\hline
$\Delta\kappa_3$ & $-\frac{1}{\lambda}C_H +3C_{H\square} -\frac{3}{4}C_{HD} -\frac{g^2}{4}(C_{HJW}-C_{2JW})$ \\
$\DkapFb$ & $-C_y +C_{H\square} -\frac{1}{4}C_{HD} -\frac{g^2}{4}(C_{HJW}-C_{2JW})$ \\
$\DkapVb$ & $C_{H\square} -\frac{1}{4}C_{HD} -\frac{3g^2}{4}(C_{HJW}-C_{2JW})$ \\
\hline
$f_{gg}$ & $\frac{4}{g_s^2}C_{HG}$ \\
$f_{z\gamma}$ & $\frac{2}{gg'}\bigl[2\cw\sw(C_{HW}-C_{HB}) -(\cw^2-\sw^2)C_{HWB}\bigr]$ \\
$f_{\gamma\gamma}$ & $4\bigl(\frac{1}{g^2}C_{HW} +\frac{1}{g'^2}C_{HB} -\frac{1}{gg'}C_{HWB}\bigr)$ \\
\hline
$c_{2y}$ & $C_{2y}$ \\
\hline
\end{tabular}
\caption{\label{tab:up} Expression of the 16 universal parameters, defined from the effective Lagrangian as in \eqref{Lu-up}, in terms of the Warsaw basis Wilson coefficients in \eqref{Lu}. These parameters completely characterize the indirect BSM effects in universal theories at the dimension-6 level. More details of the universal parameters, including their expressions in other bases, can be found in~\cite{Wells:2015uba}.}
\end{table}

If we restrict ourselves to the 16-dimensional parameter space of universal theories, a subspace of the full SMEFT parameter space, there is a unique well-motivated procedure to define the oblique parameters at LO. The field-redefinition ambiguity associated with the vector boson self-energies is eliminated by ensuring the oblique parameters defining conditions are satisfied~\cite{Barbieri:2004qk,Wells:2015uba}. At the dimension-6 level, there are 5 nonvanishing oblique parameters $\hat S$, $\hat T$, $W$, $Y$, $Z$, which constitute a subset of the 16 independent {\it universal parameters}. By our choice in~\cite{Wells:2015uba}, the latter also include: 4 anomalous triple-gauge couplings (TGCs) $\Dgzb$, $\Dkapab$, $\lamab$, $\lamgb$; 3 rescaling factors for the $h^3$, $hff$, $hVV$ vertices $\Delta\kappa_3$, $\DkapFb$, $\DkapVb$; 3 parameters for $hV_{\mu\nu}V'^{\mu\nu}$-type interactions absent in the SM $f_{gg}$, $f_{z\gamma}$, $f_{\gamma\gamma}$; 1 four-fermion coupling $c_{2y}\sim\O(y_f^2)$. As summarized in appendix~\ref{app:up}, each of these universal parameters can be identified as the coefficient of a term in $\L_\text{universal}$ in the electroweak symmetry broken phase in the unitary gauge, after the field and parameter redefinitions detailed in~\cite{Wells:2015uba}. The 16 universal parameters are just a phenomenologically convenient linear mapping from the 16 independent Wilson coefficients in~\eqref{Lu}; see table~\ref{tab:up}. As such, they constitute a complete characterization of universal theories in the SMEFT framework at the dimension-6 level.

\begin{table}[tbp]
\centering
\begin{tabular}{|l|l|}
\hline
Higgs basis coupling & Universal parameters expression \\
\hline
$\delta_m$ & $\frac{\cw^2}{\cw^2-\sw^2}\frac{\De_1}{2} -\frac{\De_2}{2} -\frac{\sw^2}{\cw^2-\sw^2}\De_3$ \\
\hline
$[\dgL^{Wf}]_{ij}\,\, (f=q,l)$ & $\delta_{ij}\bigl(\frac{\cw^2}{\cw^2-\sw^2}\frac{\De_1}{2} -\frac{\sw^2}{\cw^2-\sw^2}\De_3\bigr)$ \\
$[\dgL^{Zf}]_{ij}\,\, (f=u_L,d_L,e_L,\nu)$ & $\delta_{ij}\Bigl[T^3_f\frac{\De_1}{2} +Q_f\frac{\sw^2}{\cw^2-\sw^2}\bigl(\frac{\De_1}{2}-\De_3\bigr)\Bigr]$ \\
$[\dgR^{Zf}]_{ij}\,\, (f=u_R,d_R,e_R)$ & $\delta_{ij}Q_f\frac{\sw^2}{\cw^2-\sw^2}\bigl(\frac{\De_1}{2}-\De_3\bigr)$ \\
\hline
$\dgz$ & $\Dgzb -\frac{\De_2}{\cw^2} +\frac{\sw^2}{\cw^2-\sw^2}\bigl(\frac{\De_1}{2\sw^2}-\frac{\De_3}{\cw^2}\bigr)$ \\
$\dkapa$ & $\Dkapab$ \\
$\lambda_\gamma$ & $\lamab$ \\
$c_{3G}$ & $-\frac{2}{3g_s^2g^2}\lamgb$ \\
\hline
$\delta\lambda_3$ & $\lambda\Delta\kappa_3$ \\
$[\delta y_{f'}]_{ij}\,\, (f'=u,d,e)$ & $\delta_{ij}\DkapFb$ \\
$\delta c_z$ & $\DkapVb$ \\
\hline
$c_{gg}, c_{z\gamma}, c_{\gamma\gamma}$ & $f_{gg}, f_{z\gamma}, f_{\gamma\gamma}$, respectively \\
\hline
$c_{4f}$ & combinations of $W, Y, Z, c_{2y}$ \\
\hline
$[\dgR^{Wq}]_{ij}, [d_{Vf}]_{ij}$ & 0 \\
\hline
\end{tabular}
\caption{\label{tab:hb} Higgs basis couplings in terms of the universal parameters, taken from~\cite{Wells:2015uba}. $\De_{1,2,3}$ are independent linear combinations of $\hat S, \hat T, W, Y$ defined in \eqref{Dedef}. $c_{4f}$ collectively denotes four-fermion effective couplings, and $d_{Vf}$ stands for the dipole-type $Vff$ couplings. Compared with~\cite{HiggsBasis}, we have written the fractional $W$ mass shift as $\delta_m$ instead of $\delta m$, and defined $[\dgL^{Wq}]_{ij}$ in the gauge-eigenstate rather than mass-eigenstate basis.}
\end{table}

As yet another equivalent description of the universal theories EFT, the Higgs basis couplings, defined in~\cite{HiggsBasis} at LO in $\frac{v^2}{\Lambda^2}$, make the leading BSM effects on precision observables manifest. As ensured by the Higgs basis defining conditions~\cite{HiggsBasis,Wells:2015uba}, they capture vertex corrections involving the physical particles. Furthermore, since the input observables are not shifted at tree level, simple LO relations can be written down between some precision observables and the Higgs basis couplings. For example, the fractional shifts in $\Gamma(Z\to b_L\bar b_L)$ and $\Gamma(Z\to b_R\bar b_R)$ are proportional to $[\dgL^{Zd}]_{33}$ and $[\dgR^{Zd}]_{33}$, respectively. In general, the Higgs basis couplings are linear combinations of Wilson coefficients in the Warsaw basis (or any other complete nonredundant basis), some of which are listed in appendix~\ref{app:hb}. In the special case of universal theories, we have worked out in~\cite{Wells:2015uba} the Higgs basis couplings in terms of the universal parameters. They are reproduced here in table~\ref{tab:hb}, where the $\De$ parameters~\cite{Barbieri:2004qk,Altarelli:1990zd,Altarelli:1991fk} are 3 independent linear combinations of $\hat S$, $\hat T$, $W$, $Y$,
\beq
\De_1\equiv \hat T-W-\frac{\sw^2}{\cw^2}Y,\quad \De_2\equiv -W,\quad \De_3\equiv \hat S-W-Y.
\eeq{Dedef}
A universal pattern of fermion couplings can be seen from table~\ref{tab:hb}. In particular, all the $Vff$ vertex corrections depend on just 2 parameters $\De_1$, $\De_3$, and all the $hff$ vertices are rescaled by a common factor $(1+\DkapFb)$. This is not the case for generic nonuniversal theories, where the number of independent couplings is equal to the number of independent dimension-6 operators in the full SMEFT. For universal theories, on the other hand, the generically independent couplings are related as follows,
\bseq
\beqa
&& \dgL^{Wq} = \dgL^{Wl},\quad \frac{\dgR^{Zu}}{Y_u} = \frac{\dgR^{Zd}}{Y_d} =  \frac{\dgR^{Ze}}{Y_e}, \CR
&& \dgL^{Ze}+\dgL^{Z\nu} = \dgR^{Ze},\quad \dgL^{Zu}+\dgL^{Zd} = \dgR^{Zu}+\dgR^{Zd}, \label{ur-ew}\\
&& \delta y_u = \delta y_d = \delta y_e = \DkapFb.\label{ur-yuk}
\eeqan
\eseq{ur}
We will call \eqref{ur} ``{\it universal relations}'' from here on. Compared with~\cite{Wells:2015uba}, we have replaced $Q_u$, $Q_d$, $Q_e$ by the equivalent $Y_u$, $Y_d$, $Y_e$ for later convenience. Each Higgs basis coupling appearing in \eqref{ur} represents the diagonal elements of a $3\times3$ matrix in generation space that is proportional to $\delta_{ij}$ for universal theories. Additional universal relations among 4-fermion couplings can be written down, which do not concern us here. Essentially, the universal relations among the generically independent Higgs basis couplings are in exact correspondence with the correlations among the otherwise independent fermionic operator Wilson coefficients shown in table~\ref{tab:Qf}, e.g.\
\beq
\dgL^{Wq} = \dgL^{Wl} \quad\Leftrightarrow\quad [C_{Hq}^{(3)}]_{ij} = [C_{Hl}^{(3)}]_{ij} \,\,\Bigl(=\delta_{ij}\frac{g^2}{4}C_{HJW}\Bigr).
\eeqn

%%%
\subsection{Overview of RG-induced nonuniversal effects}
\label{sec:nu}

Beyond LO, renormalization is needed, and the Wilson coefficients as renormalized Lagrangian parameters should have renormalization scales $\mu$ associated with them. The scale dependence of the Wilson coefficients is captured by the RG equations, which at leading order are governed by the anomalous dimensions $\gamma_{ij}$,
\beq
\dot C_i \equiv 16\pi^2\frac{d}{d\ln\mu}C_i(\mu) = \sum_j \gamma_{ij}C_j(\mu).
\eeq{RGeq}
It should be emphasized that $\gamma_{ij}$ are unambiguous only when a complete nonredendant basis of effective operators is specified. The Warsaw basis adopted here is the same basis used in~\cite{Jenkins:2013zja,Jenkins:2013wua,Alonso:2013hga} to calculate the full $\gamma_{ij}$ matrix for the dimension-6 operators.

The renormalization scale $\mu$ should be properly chosen to avoid large radiative corrections beyond a fixed-order calculation. If we are interested in the deviations of precision observables at the electroweak scale, $\mu\sim\muew$ is desired, because large logarithms in the perturbative expansion can be avoided when the observables are expressed in terms of $C_i(\muew)$. But on the other hand, to infer the shape of the UV theory at a higher scale $\Lambda\gg\muew$, which is the ultimate goal of SMEFT analyses, $C_i(\Lambda)$ are needed, because we should better set $\mu\sim\Lambda$ when the SMEFT is matched onto a specific new physics model in the UV. Solving \eqref{RGeq} to leading order, which is sufficient for most practical purposes, we obtain
\beq
C_i(\muew) = C_i(\Lambda) -\frac{1}{16\pi^2}\ln\frac{\Lambda}{\muew} \sum_j \gamma_{ij}C_j(\Lambda).
\eeq{RGsol}
The second term in this equation contributes to the LL corrections of the observables which are affected by $C_iQ_i$ at LO, when they are calculated in terms of the Wilson coefficients at $\mu=\Lambda$. To be specific, up to higher-order terms, the fractional shift of an observable $\Obs$ is
\beq
\dbNP\Obs \equiv \frac{\Obs-\Obs^\SM}{\Obs^\SM} = \sum_i a_i C_i(\muew) = \sum_i a_i C_i(\Lambda) -\frac{1}{16\pi^2}\ln\frac{\Lambda}{\muew}\sum_{i,j}a_i\gamma_{ij}C_j(\Lambda),
\eeq{dbNPO}
where $a_i$ are functions of properly-renormalized SM parameters, which %are appropriately renormalized around $\muew$, and 
can be recast in terms of the input observables~\cite{Wells:2014pga}. It is based on \eqref{dbNPO} that constraints on $C_i(\muew)$ derived from precision data can be translated into constraints on (combinations of) $C_j(\Lambda)$'s, some of which are less accessible otherwise; see e.g.~\cite{Mebane:2013zga,Elias-Miro:2013mua,Elias-Miro:2013eta,deBlas:2015aea}.

For universal theories, a key observation is that the correlations among the fermionic operator Wilson coefficients at the matching scale $\Lambda$, represented by a set of linear equations
\beq
\sum_i b_i C_i(\Lambda) = 0,
\eeqn
are not necessarily preserved by RG evolution, because it is possible that
\beq
\sum_{i,j} b_i \gamma_{ij} C_j(\Lambda) \ne 0.
\eeqn
As a consequence, at the electroweak scale $\muew$ where precision observables are measured, we may have
\beq
\sum_i b_i C_i(\muew) \ne 0.
\eeq{nuC}
For example, while $[C_{Hq}^{(3)}]_{ij} -[C_{Hl}^{(3)}]_{ij}=0$ at $\mu=\Lambda$ for universal theories, the same is in general not true at $\mu=\muew$, as we will show in section~\ref{sec:ew}. When \eqref{nuC} happens, the universal theory at $\Lambda$ flows to a nonuniversal theory at $\muew$. We say that nonuniversal effects are induced by RG evolution.

The observation above poses a challenge for defining oblique parameters, and more generally universal parameters, in the SMEFT at the electroweak scale. In general, the oblique parameters defining conditions, which require the absence of fermionic operators, cannot be satisfied no matter how the fields and parameters are redefined due to the theory being nonuniversal. Additional prescriptions are needed if one wishes to define and use these parameters, which can be somewhat arbitrary. This also means that without additional prescriptions, it is in general not meaningful to talk about RG evolution of the universal parameters.

Nevertheless, as far as observables are concerned, there are no ambiguities, since \eqref{dbNPO}, which relates $\dbNP\Obs$ to $C_i(\Lambda)$ at LL accuracy, always holds. With the linear mapping in table~\ref{tab:up}, we can recast $\dbNP\Obs$ in terms of the 16 universal parameters defined at the matching scale, $\hat S(\Lambda)$, $\hat T(\Lambda)$, etc., as long as the theory is universal at $\Lambda$. The RG-induced nonuniversal effects then manifest themselves in the fact that all the LL corrections in \eqref{dbNPO} cannot be absorbed into the running of the parameters appearing in the LO expression for $\dbNP\Obs$, namely the 16 universal parameters. In the following sections, we will define $\hat S(\muew)$, $\hat T(\muew)$, etc.\ to absorb part of the LL corrections, following some well-motivated additional prescriptions. The prediction for $\dbNP\Obs$ then involves the LO expression in terms of these universal parameters at $\muew$, plus additional LL terms. The presence of the latter may potentially affect the interpretation and usefulness of global fits to observables at $\muew$ assuming the theory is universal at this scale, including the conventional oblique parameters fits. But when they are taken into account, consistent constraints on $\hat S(\Lambda)$, $\hat T(\Lambda)$, etc.\ at the LL level can in principle be derived from precision data, which can be used to infer the BSM new physics at $\Lambda$ if it is universal.\footnote{The accuracy of the LL-level constraints on the universal parameters is a separate issue that deserves further investigation. %The NLO finite terms not enhanced by $\ln\frac{\Lambda}{\muew}$ may be important for some observables~\cite{Hartmann:2015oia,Hartmann:2015aia}. 
If the LL corrections are important for some observables, the NLO finite terms not enhanced by $\ln\frac{\Lambda}{\muew}$ may also be~\cite{Hartmann:2015oia,Hartmann:2015aia}. In any case, the effect of the neglected terms in a finite-order perturbative calculation may be accounted for by introducing SMEFT theory uncertainties, as advocated recently in~\cite{Berthier:2015oma,Berthier:2015gja} in the more general context of fitting the full SMEFT.} 

The close connection between the Higgs basis couplings and precision observables at LO offers an equivalent and convenient way to formulate the analysis. While it is still a matter of debate how to extend the Higgs basis framework beyond LO, at least at the LL level there is a straightforward procedure. In the full SMEFT at the dimension-6 level, we can think of the Higgs basis couplings as {\it defined by} the linear combinations of Wilson coefficients in the Warsaw basis (or any other complete nonredundant basis) worked out in~\cite{HiggsBasis}, {\it with the renormalization scale dependence included}. For example, in our notation,
\bseq
\beqa
[\dgL^{Wl}(\mu)]_{ij} &\equiv& [C_{Hl}^{(3)}(\mu)]_{ij} -\frac{\cw\sw}{\cw^2-\sw^2} C_{HWB}(\mu) -\frac{\cw^2}{\cw^2-\sw^2}C_0(\mu),\label{dgLWlmu} \\
{[}\dgL^{Wq}(\mu)]_{ij} &\equiv& [C_{Hq}^{(3)}(\mu)]_{ij} -\frac{\cw\sw}{\cw^2-\sw^2} C_{HWB}(\mu) -\frac{\cw^2}{\cw^2-\sw^2}C_0(\mu),\label{dgLWqmu} 
%{[}\dgL^{Zd}(\mu)]_{ij} &\equiv& -\frac{1}{2}[C_{Hq}^{(3)}(\mu)]_{ij} -\frac{1}{2}[C_{Hq}^{(1)}(\mu)]_{ij} +\frac{\cw\sw}{3(\cw^2-\sw^2)} C_{HWB}(\mu) +\frac{3\cw^2-\sw^2}{6(\cw^2-\sw^2)}C_0(\mu),\CR \\
%{[}\dgR^{Zd}(\mu)]_{ij} &\equiv& -\frac{1}{2}[C_{Hd}(\mu)]_{ij} +\frac{\cw\sw}{3(\cw^2-\sw^2)} C_{HWB}(\mu) +\frac{\sw^2}{3(\cw^2-\sw^2)}C_0(\mu),
\eeqan
\eseq{dgLWfmu}
where
\beq
C_{0}(\mu) \equiv \frac{1}{4} \Bigl\{C_{HD}(\mu) +2 \bigl([C_{Hl}^{(3)}(\mu)]_{11} +[C_{Hl}^{(3)}(\mu)]_{22}\bigr) -\bigl([C_{ll}(\mu)]_{1221}+[C_{ll}(\mu)]_{2112}\bigr)\Bigr\}
\eeq{C0def}
is a combination of Wilson coefficients coming from undoing the shifts in the input observables $m_Z$ and $G_F$, as required by the Higgs basis defining conditions. Note that $\cw=\frac{g}{\sqrt{g^2+g'^2}}$, $\sw=\frac{g'}{\sqrt{g^2+g'^2}}$ are also $\mu$-dependent. The running of the Higgs basis couplings with $\mu$ follows from the RG equations for the Warsaw basis Wilson coefficients and the SM parameters. For universal theories at $\Lambda$, the universal relations in \eqref{ur} should actually read $\dgL^{Wq}(\Lambda) = \dgL^{Wl}(\Lambda)$, etc. After RG evolution down to the electroweak scale, these relations are violated in the sense that $\dgL^{Wq}(\muew) \ne \dgL^{Wl}(\muew)$, etc., due to $[C_{Hq}^{(3)}(\muew)]_{ij} \ne [C_{Hl}^{(3)}(\muew)]_{ij}$, etc., as mentioned below \eqref{nuC}. This was already alluded to in figure~\ref{fig:diag}, and will be demonstrated in detail in the next section. Defined in this way, the Higgs basis couplings renormalized at $\muew$ directly map onto $\dbNP\Obs$. Two example observables we will discuss later are $\Rl\equiv\Gamhad/\Gamma(Z\to\ell^+\ell^-)$ (assuming lepton flavor universality) and $\Rb\equiv\Gamma(Z\to b\bar b)/\Gamhad$, where $\Gamhad$ is the hadronic $Z$ decay partial width. From their LO expressions,
\bseq
\beqa
\Rl &=& \frac{3\Bigl\{\sum\limits_{i=1}^{2}\Bigl[\bigl([g_L^{Zu}]_{ii}\bigr)^2+\bigl([g_R^{Zu}]_{ii}\bigr)^2\Bigr] +\sum\limits_{i=1}^{3}\Bigl[\bigl([g_L^{Zd}]_{ii}\bigr)^2+\bigl([g_R^{Zd}]_{ii}\bigr)^2\Bigr]\Bigr\}}{\bigl([g_L^{Ze}]_{jj}\bigr)^2+\bigl([g_R^{Ze}]_{jj}\bigr)^2}\quad (j=1,2,\,\text{or}~3),\CR \\
\Rb &=& \frac{\bigl([g_L^{Zd}]_{33}\bigr)^2+\bigl([g_R^{Zd}]_{33}\bigr)^2}{\sum\limits_{i=1}^{2}\Bigl[\bigl([g_L^{Zu}]_{ii}\bigr)^2+\bigl([g_R^{Zu}]_{ii}\bigr)^2\Bigr] +\sum\limits_{i=1}^{3}\Bigl[\bigl([g_L^{Zd}]_{ii}\bigr)^2+\bigl([g_R^{Zd}]_{ii}\bigr)^2\Bigr]}, %\\
\eeqan
\eseqn
where $[g_{L,R}^{Zf}]_{ij}=\delta_{ij}g_{L,R}^{Zf}+[\delta g_{L,R}^{Zf}(\muew)]_{ij}$ with $g_L^{Zf}\equiv T^3_f-Q_f\sw^2(\muew)$, $g_R^{Zf}\equiv-Q_f\sw^2(\muew)$, it follows that the fractional corrections with respect to the SM are given by
\bseq
\beqa
\hspace{-4ex}\dbNP\Rl &=& \dbNP\Gamhad -\frac{2}{(g_L^{Ze})^2+(g_R^{Ze})^2}\bigl(g_L^{Ze}[\dgL^{Ze}(\muew)]_{jj} +g_R^{Ze}[\dgR^{Ze}(\muew)]_{jj}\bigr), \\
\hspace{-4ex}\dbNP\Rb &=& \frac{2}{(g_L^{Zd})^2+(g_R^{Zd})^2}\bigl(g_L^{Zd}[\dgL^{Zd}(\muew)]_{33} +g_R^{Zd}[\dgR^{Zd}(\muew)]_{33}\bigr) -\dbNP\Gamhad,
\eeqan
\eseq{dbNPRlRb}
where
\beqa
\dbNP\Gamhad &=& \frac{2}{2\Bigl[(g_L^{Zu})^2+(g_R^{Zu})^2\Bigr] +3\Bigl[(g_L^{Zd})^2+(g_R^{Zd})^2\Bigr]}\times \CR
&&\quad \biggl\{ \sum_{i=1}^{2}\Bigl(g_L^{Zu}[\dgL^{Zu}(\muew)]_{ii}+g_R^{Zu}[\dgR^{Zu}(\muew)]_{ii}\Bigr) \CR
&&\quad\,\, +\sum_{i=1}^{3}\Bigl(g_L^{Zd}[\dgL^{Zd}(\muew)]_{ii}+g_R^{Zd}[\dgR^{Zd}(\muew)]_{ii}\Bigr) \biggr\}.
\eeqan
The Higgs basis couplings involved in these equations are given by \eqref{dgZf} in terms of the Warsaw basis Wilson coefficients.

To end this subsection, we comment on a subtlety associated with defining phenomenological parameters in the electroweak symmetry broken phase. The renormalized vacuum expectation value of the Higgs field %, denoted by $v_r(\mu)$,
 is a scheme-dependent quantity. To avoid introducing unnecessary scheme dependence into the running of the Wilson coefficients, we take the $v$ appearing in \eqref{Lu} to be simply a constant, say 246.2~GeV. %As a consequence, factors of $\frac{2|\langle H\rangle|^2}{v^2}$ appear in the universal parameters and Higgs basis couplings
As a consequence, when the universal parameters and Higgs basis couplings, defined from the effective Lagrangian in the broken phase, are calculated in terms of the Wilson coefficients, factors of $\frac{2|\langle H\rangle|^2}{v^2}=1+\dots$ appear. We treat the $\dots$ pieces as part of the one-loop counterterms, not to be included in the renormalized Higgs basis couplings, or renormalized universal parameters when they are properly defined. These terms are relevant for a full NLO calculation, but do not affect the LL corrections proportional to $\ln\frac{\Lambda}{\muew}$ that are the focus of the present paper.

%%%
\section{RG effects in the electroweak sector}
\label{sec:ew}

%%%
\subsection{The universal limit}

We first look at the electroweak sector, and begin with the limit $y_f\to0$. The Lagrangian at the new physics scale $\Lambda$ is \eqref{Lu} with $C_y=C_{2y}=0$. We see from table~\ref{tab:Qf} that the $\psi^2H^2D$-class operators, which affect the $Vff$ effective couplings, are related in universal theories at LO as follows,
\bseq
\beqa
&& [C_{Hq}^{(3)}]_{ij} = [C_{Hl}^{(3)}]_{ij} = \delta_{ij}\frac{g^2}{4}C_{HJW}, \\
&& \bigl[ \{C_{Hq}^{(1)}, C_{Hl}^{(1)}, C_{Hu}, C_{Hd}, C_{He}\} \bigr]_{ij} = \{ Y_q, Y_l, Y_u, Y_d, Y_e \} \delta_{ij} \frac{g'^2}{2} C_{HJB}.
\eeqan
\eseq{ur-ew-C}
These relations are equivalent to the universal relations in \eqref{ur-ew}. Using the formulas in~\cite{Alonso:2013hga}, we find the one-loop running of these Wilson coefficients,
\bseq
\beqa
[\dot C_{Hq}^{(3)}]_{ij} &=& \delta_{ij}g^2 \Bigl(\frac{1}{6}C_{H\square} +\frac{7}{12}g^2C_{HJW} +\frac{23}{6}g^2C_{2JW} +\frac{1}{54}g'^2C_{2JB} +\frac{8}{9}g_s^2C_{2JG} \Bigr), \\
{[}\dot C_{Hl}^{(3)}]_{ij} &=& \delta_{ij}g^2 \Bigl(\frac{1}{6}C_{H\square} +\frac{7}{12}g^2C_{HJW} +\frac{23}{6}g^2C_{2JW} +\frac{1}{6}g'^2C_{2JB} \Bigr), \\
{[}\dot C_{Hq}^{(1)}]_{ij} &=& Y_q \delta_{ij}g'^2 \Bigl[\frac{1}{3}(C_{H\square}+C_{HD}) +\frac{41}{6}g'^2C_{HJB} \CR
&&\qquad\qquad +g^2C_{2JW} +\frac{361}{27}g'^2C_{2JB} +\frac{16}{9}g_s^2C_{2JG} \Bigr], \\
{[}\dot C_{Hl}^{(1)}]_{ij} &=& Y_l \delta_{ij}g'^2 \Bigl[\frac{1}{3}(C_{H\square}+C_{HD}) +\frac{41}{6}g'^2C_{HJB} +g^2C_{2JW} +\frac{41}{3}g'^2C_{2JB} \Bigr], \\
{[}\dot C_{Hu}]_{ij} &=& Y_u \delta_{ij}g'^2 \Bigl[\frac{1}{3}(C_{H\square}+C_{HD}) +\frac{41}{6}g'^2C_{HJB} +\frac{376}{27}g'^2C_{2JB} +\frac{16}{9}g_s^2C_{2JG} \Bigr], \\
{[}\dot C_{Hd}]_{ij} &=& Y_d \delta_{ij}g'^2 \Bigl[\frac{1}{3}(C_{H\square}+C_{HD}) +\frac{41}{6}g'^2C_{HJB} +\frac{364}{27}g'^2C_{2JB} +\frac{16}{9}g_s^2C_{2JG} \Bigr], \\
{[}\dot C_{He}]_{ij} &=& Y_e \delta_{ij}g'^2 \Bigl[\frac{1}{3}(C_{H\square}+C_{HD}) +\frac{41}{6}g'^2C_{HJB} +\frac{44}{3}g'^2C_{2JB} \Bigr].
\eeqan
\eseq{CHfrun}
Note that only the Wilson coefficients that are nonzero at LO (i.e.\ at $\mu=\Lambda$) need to be kept on the RHS of these equations. We have used table~\ref{tab:Qf}, or equivalently \eqref{ur-ew-C} and \eqref{ur-4f-C} below, to rewrite them in terms of the coefficients of the operator combinations in \eqref{Lu} for universal theories.

From the discussion in section~\ref{sec:nu}, it is clear that the relations in \eqref{ur-ew-C} are preserved by RG evolution only in the limit $C_{2JW}=C_{2JB}=C_{2JG}=0$, namely $W=Y=Z=0$ at LO (i.e.\ at $\mu=\Lambda$). We call this limit, together with $y_f\to0$, {\it the ``universal limit.''}

In the universal limit, fermionic operators in the electroweak sector are generated by RG evolution, but they are organized into the combinations $Q_{HJW}$, $Q_{HJB}$ that appear in the LO Lagrangian for universal theories. Thus, without any further prescriptions, it is unambiguous to define $C_{HJW}, C_{HJB}$ at the electroweak scale, and write down their RG equations,
\bseq
\beqa
\dot C_{HJW} &=& \frac{2}{3}C_{H\square} +\frac{26}{3}g^2C_{HJW}, \\
\dot C_{HJB} &=& \frac{2}{3}(C_{H\square}+C_{HD}).
\eeqan
\eseq{CHJVrun}
These are derived from
\bseq
\beqa
&& 16\pi^2\ddlm(g^2C_{HJW}) %= 2g\dot gC_{HJW} +g^2\dot C_{HJW} 
= g^2\Bigl(\frac{2}{3}C_{H\square} +\frac{7}{3}g^2C_{HJW}\Bigr), \\
&& 16\pi^2\ddlm(g'^2C_{HJB}) %= 2g'\dot gC_{HJB} +g'^2\dot C_{HJB} 
= g'^2\Bigl[\frac{2}{3}(C_{H\square}+C_{HD}) +\frac{41}{3}g'^2C_{HJB}\Bigr],
\eeqan
\eseqn
which follow from \eqref{ur-ew-C}, \eqref{CHfrun}, and the well-known one-loop running of the SM gauge couplings
\beq
\dot g = -\frac{19}{6}g^3,\quad \dot g' = \frac{41}{6}g'^3.
\eeq{gSMrun}

We can extend this analysis to the 4-fermion interactions. The correlations among the Wilson coefficients, i.e.\ the counterparts of \eqref{ur-ew-C}, can be read off from table~\ref{tab:Qf} (see also~\cite{Falkowski:2015krw}), with contributions from $Q_{2y}$ neglected for the moment,
\bseq
\beqa
&& [C_{ll}]_{ijkl} = \Bigl(\frac{1}{2}\delta_{il}\delta_{jk} -\frac{1}{4}\delta_{ij}\delta_{kl}\Bigr)g^2 C_{2JW} +Y_l^2\delta_{ij}\delta_{kl}g'^2 C_{2JB}, \\
&& [C_{qq}^{(3)}]_{ijkl} = \frac{1}{4}\delta_{ij}\delta_{kl}g^2 C_{2JW} +\frac{1}{4}\delta_{il}\delta_{jk}g_s^2 C_{2JG}, \\
&& [C_{lq}^{(3)}]_{ijkl} = \frac{1}{2}\delta_{ij}\delta_{kl}g^2 C_{2JW}, \\
&& \bigl[\{C_{lq}^{(1)}, C_{ee}, C_{ud}^{(1)}, C_{eu}, C_{ed}, C_{qu}^{(1)}, C_{qd}^{(1)}, C_{qe}, C_{lu}, C_{ld}, C_{le}\}\bigr]_{ijkl} = \CR
&&\quad \{2Y_lY_q, Y_e^2, 2Y_uY_d, 2Y_uY_e, 2Y_dY_e, 2Y_qY_u, 2Y_qY_d, 2Y_qY_e, 2Y_lY_u, 2Y_lY_d, 2Y_lY_e\}\CR
&&\qquad \delta_{ij}\delta_{kl}g'^2 C_{2JB}, \\
&& [C_{qq}^{(1)}]_{ijkl} = Y_q^2\delta_{ij}\delta_{kl}g'^2 C_{2JB} +\Bigl(\frac{1}{4}\delta_{il}\delta_{jk} -\frac{1}{6}\delta_{ij}\delta_{kl}\Bigr)g_s^2 C_{2JG}, \\
&& \bigl[\{C_{uu}, C_{dd}\}\bigr]_{ijkl} = \{Y_u^2, Y_d^2\}\delta_{ij}\delta_{kl}g'^2 C_{2JB} +\Bigl(\frac{1}{2}\delta_{il}\delta_{jk} -\frac{1}{6}\delta_{ij}\delta_{kl}\Bigr)g_s^2 C_{2JG}, \\
&& [C_{ud}^{(8)}]_{ijkl} = [C_{qu}^{(8)}]_{ijkl} = [C_{qd}^{(8)}]_{ijkl} = 2\delta_{ij}\delta_{kl} g_s^2 C_{2JG}.
\eeqan
\eseq{ur-4f-C}
For $C_{2JW}=C_{2JB}=C_{2JG}=0$ at LO, we find, from~\cite{Alonso:2013hga},
\bseq
\beqa
&& [\dot C_{ll}]_{ijkl} = \Bigl(\frac{1}{2}\delta_{il}\delta_{jk} -\frac{1}{4}\delta_{ij}\delta_{kl}\Bigr)g^2 \Bigl(\frac{g^2}{3}C_{HJW}\Bigr) +Y_l^2\delta_{ij}\delta_{kl}g'^2\Bigl(\frac{g'^2}{3}C_{HJB}\Bigr), \\
&& \bigl[\{\dot C_{qq}^{(3)}, \dot C_{lq}^{(3)}\}\bigr]_{ijkl} = \Bigl\{\frac{1}{4}, \frac{1}{2}\Bigr\}\delta_{ij}\delta_{kl}g^2 \Bigl(\frac{g^2}{3}C_{HJW}\Bigr), \\
&& \bigl[\{\dot C_{qq}^{(1)}, \dot C_{lq}^{(1)}, \dot C_{uu}, \dot C_{dd}, \dot C_{ee}, \dot C_{ud}^{(1)}, \dot C_{eu}, \dot C_{ed}, \dot C_{qu}^{(1)}, \dot C_{qd}^{(1)}, \dot C_{qe}, \dot C_{lu}, \dot C_{ld}, \dot C_{le} \}\bigr]_{ijkl} = \CR
&&\quad \{Y_q^2, 2Y_lY_q, Y_u^2, Y_d^2, Y_e^2, 2Y_uY_d, 2Y_uY_e, 2Y_dY_e, 2Y_qY_u, 2Y_qY_d, 2Y_qY_e, 2Y_lY_u, 2Y_lY_d, 2Y_lY_e\}\CR
&&\qquad \delta_{ij}\delta_{kl}g'^2\Bigl(\frac{g'^2}{3}C_{HJB}\Bigr), \\
&& [\dot C_{ud}^{(8)}]_{ijkl} = [\dot C_{qu}^{(8)}]_{ijkl} = [\dot C_{qd}^{(8)}]_{ijkl} = 0.
\eeqan
\eseqn
The pattern in these equations, when compared with \eqref{ur-4f-C}, indicates that in the universal limit defined above, the 4-fermion interactions are also universal after RG evolution. Thus, as in \eqref{CHJVrun}, we can unambiguously define
\bseq
\beqa
C_{2JW}(\muew) &=& -\frac{1}{16\pi^2}\ln\frac{\Lambda}{\muew} \dot C_{2JW}, \\
C_{2JB}(\muew) &=& -\frac{1}{16\pi^2}\ln\frac{\Lambda}{\muew} \dot C_{2JB},
\eeqan
\eseqn
where
\bseq
\beqa
\dot C_{2JW} &=& \frac{g^2}{3}C_{HJW}, \\
\dot C_{2JB} &=& \frac{g'^2}{3}C_{HJB}.
\eeqan
\eseq{C2JVrun}
Here the running of $g$ and $g'$ is not relevant, since $\dot g, \dot g'$ are multiplied by the values of $C_{2JW}, C_{2JB}$ at LO which vanish. We see that, if the operators $Q_{2JW}, Q_{2JB}$ are not generated by the universal new physics at $\mu=\Lambda$, they will be generated at one-loop level by RG evolution down to $\mu=\muew$, and result in a universal pattern in the 4-fermion interactions at the electroweak scale. The operator $Q_{2JG}$, on the other hand, is not generated by RG evolution at this order if it is absent at the new physics scale.

Eqs. \eqref{CHJVrun} and \eqref{C2JVrun} allow us to write down meaningful RG equations for the oblique parameters in the universal limit, namely $y_f=0$, and $C_{2JW}=C_{2JB}=C_{2JG}=0$, or equivalently $W=Y=Z=0$, at $\mu=\Lambda$. To do so, we further need table~\ref{tab:up}, the RG equations for the bosonic Wilson coefficients from~\cite{Jenkins:2013zja,Alonso:2013hga},
\bseq
\beqa
\dot C_{HWB} &=& \Bigl(\frac{4}{3}g^2 +\frac{19}{3}g'^2 +4\lambda\Bigr)C_{HWB} -3g^2g'C_W +2gg'(C_{HW}+C_{HB}), \\
\dot C_{HD} &=& \Bigl(\frac{9}{2}g^2 -\frac{5}{6}g'^2 +12\lambda \Bigr) C_{HD} +\frac{20}{3}g'^2C_{H\square} +\frac{40}{3}g'^4C_{HJB},
\eeqan
\eseq{Cbosrun}
and the running of the SM gauge couplings \eqref{gSMrun}. The results are
\bseq
\beqa
\dot{\hat S} &=& -\frac{1}{3}(19g^2 -g'^2)\hat S -\frac{1}{2}g^2\hat T -\frac{1}{3}(27g^2 -g'^2)\cw^2\Dgzb +\frac{1}{6}(33g^2 +g'^2+24\lambda)\Dkapab +2g^2\lamab \CR
&& +\frac{1}{3}g^2\DkapVb +\frac{1}{2}g^2(g^2-g'^2)f_{z\gamma} +e^2g^2f_{\gamma\gamma}\\
\dot{\hat T} &=& \frac{3}{2}(3g^2+8\lambda)\Bigl[\hat T -2\frac{\sw^2}{\cw^2}(\hat S-\Dkapab)\Bigr] %+12\lambda\frac{\sw^2}{\cw^2}Y 
-24\lambda\sw^2\Dgzb -3g'^2\DkapVb, \\
\dot W &=& \frac{2}{3}g^2\cw^2\Dgzb, \\
%\dot Y &=& -\frac{2}{3}g'^2\hat S -\frac{2}{3}e^2\Dgzb +\frac{2}{3}g'^2\Dkapab, \\
\dot Y &=& -\frac{2}{3}g'^2(\hat S +\cw^2\Dgzb -\Dkapab).%, \\
%\dot Z &=& 0,
\eeqan
\eseq{STWYrun}
%where we have also included $\dot Z$ for completeness. 
Similarly, $\dot Z = 0$. In \eqref{STWYrun} we have recast the Wilson coefficients on the RHS in terms of universal parameters. Following these evolution equations from $\Lambda$ to $\muew$, we obtain the oblique parameters at the electroweak scale, 
\bseq
\beqa
\hat S(\muew) &=& \hat S(\Lambda) -\frac{1}{16\pi^2}\ln\frac{\Lambda}{\muew}\dot{\hat S}, \\
\hat T(\muew) &=& \hat T(\Lambda) -\frac{1}{16\pi^2}\ln\frac{\Lambda}{\muew}\dot{\hat T}, \\
W(\muew) &=& W(\Lambda) -\frac{1}{16\pi^2}\ln\frac{\Lambda}{\muew}\dot W, \\
Y(\muew) &=& Y(\Lambda) -\frac{1}{16\pi^2}\ln\frac{\Lambda}{\muew}\dot Y,
%Z(\muew) &=& Z(\Lambda) -\frac{1}{16\pi^2}\ln\frac{\Lambda}{\muew}\dot Z,
\eeqan
\eseq{STWYewdef}
which  are to be used to calculate the observables, or alternatively, the Higgs basis couplings at $\mu=\muew$, in the electroweak sector. For example,
\beqa
\hspace{-4ex}&&[\dgL^{Wl}(\muew)]_{ij} = [\dgL^{Wq}(\muew)]_{ij} = \delta_{ij}\Bigl[\frac{\cw^2}{\cw^2-\sw^2}\frac{\De_1(\muew)}{2} -\frac{\sw^2}{\cw^2-\sw^2}\De_3(\muew)\Bigr] \CR
\hspace{-4ex}&&\quad = \frac{\delta_{ij}}{2(\cw^2-\sw^2)}\Bigl[-2\sw^2\hat S(\muew) +\cw^2\hat T(\muew) -(\cw^2-2\sw^2)W(\muew) +\sw^2Y(\muew) \Bigr],
\eeqa{dgLWfu}
where the SM parameters $\cw$, $\sw$ are also renormalized at $\mu=\muew$. We stress again that \eqref{STWYrun}, \eqref{STWYewdef}, \eqref{dgLWfu} are unambiguous only in the universal limit $W(\Lambda)=Y(\Lambda)=Z(\Lambda)=0$, $y_f=0$ [we have kept $W(\Lambda)$, $Y(\Lambda)$ in \eqref{STWYewdef} for later convenience]; otherwise the theory becomes nonuniversal after RG evolution and it is not even clear how to define the oblique parameters at $\muew$. We will go beyond this limit in the next subsection.

One interesting observation from \eqref{STWYrun} is that, with our definition of universal parameters, and in the special universal limit discussed above where these equations are meaningful, the $\hat S$ and $\hat T$ parameters mix under RG evolution. This is true despite the fact that $C_{HWB}$ and $C_{HD}$, which {\it contribute to} $\hat S$ and $\hat T$, respectively, do not mix in the Warsaw basis, even when the full SMEFT is considered~\cite{Alonso:2013hga}. The reason is that, as is clear from table~\ref{tab:up}, $\hat S$ and $\hat T$ {\it should not be identified with} $C_{HWB}$ and $C_{HD}$. The additional contributions to these oblique parameters lead to the mixing observed here.

%%%
\subsection{Nonuniversal effects beyond the universal limit}

Now we are ready to turn back on the LO $C_{2JW}, C_{2JB}, C_{2JG}$ (while still assuming $y_f\to0$), and study the nonuniversal effects due to their contributions to the RG evolution. These effects are conveniently represented by violations of the universal relations \eqref{ur-ew}. Using \eqref{CHfrun}, together with the relations between the Higgs basis couplings and the Warsaw basis Wilson coefficients given by \eqref{dgLWfmu} and \eqref{dgZf}, we find
\bseq
\beqa
&& \dgLd^{Wq} -\dgLd^{Wl} = \dot C_{Hq}^{(3)} -\dot C_{Hl}^{(3)} = g^2\Bigl(-\frac{4}{27}g'^2C_{2JB} +\frac{8}{9}g_s^2C_{2JG}\Bigr) = \frac{8}{27}(g'^2Y -6g_s^2Z), \CR\\
&& \frac{\dgRd^{Zu}}{Y_u} -\frac{\dgRd^{Zd}}{Y_d} = -\frac{1}{2}\Bigl(\frac{\dot C_{Hu}}{Y_u} -\frac{\dot C_{Hd}}{Y_d}\Bigr) = -\frac{g'^2}{2}\frac{4}{9}g'^2C_{2JB} = \frac{4}{9}\frac{\sw^2}{\cw^2}g'^2Y, \\
&& \frac{\dgRd^{Zd}}{Y_d} -\frac{\dgRd^{Ze}}{Y_e} = -\frac{1}{2}\Bigl(\frac{\dot C_{Hd}}{Y_d} -\frac{\dot C_{He}}{Y_e}\Bigr) \CR
&&\qquad = -\frac{g'^2}{2}\Bigl(-\frac{32}{27}g'^2C_{2JB} +\frac{16}{9}g_s^2C_{2JG}\Bigr) = \frac{16}{27}\frac{\sw^2}{\cw^2} (-2g'^2Y +3g_s^2Z), \\
&& \dgLd^{Ze}+\dgLd^{Z\nu}-\dgRd^{Ze} = -\frac{1}{2}(2\dot C_{Hl}^{(1)}-\dot C_{He}) \CR
&&\qquad = -\frac{g'^2}{2}(-g^2C_{2JW} +g'^2C_{2JB}) = \frac{\sw^2}{\cw^2}(-g^2W +g'^2Y), \\
&& \dgLd^{Zu}+\dgLd^{Zd}-\dgRd^{Zu}-\dgRd^{Zd} = -\frac{1}{2}(2\dot C_{Hq}^{(1)}-\dot C_{Hu}-\dot C_{Hd}) \CR
&&\qquad = -\frac{g'^2}{2}\Bigl(\frac{1}{3}g^2C_{2JW} -\frac{1}{3}g'^2C_{2JB} \Bigr) = \frac{1}{3}\frac{\sw^2}{\cw^2} (g^2W-g'^2Y),
\eeqan
\eseqn
where diagonal elements have been assumed for the matrices in generation space. It follows that at the electroweak scale,
\bseq
\beqa
&& \dgL^{Wq}(\muew) -\dgL^{Wl}(\muew) = -\frac{1}{16\pi^2}\ln\frac{\Lambda}{\muew}\cdot\frac{8}{27}(g'^2Y -6g_s^2Z), \\
&& \frac{\dgR^{Zu}(\muew)}{Y_u} -\frac{\dgR^{Zd}(\muew)}{Y_d} = -\frac{1}{16\pi^2}\ln\frac{\Lambda}{\muew}\cdot\frac{4}{9}\frac{\sw^2}{\cw^2}g'^2Y, \\
&& \frac{\dgR^{Zd}(\muew)}{Y_d} -\frac{\dgR^{Ze}(\muew)}{Y_e} = -\frac{1}{16\pi^2}\ln\frac{\Lambda}{\muew}\cdot\frac{16}{27}\frac{\sw^2}{\cw^2}(-2g'^2Y +3g_s^2Z), \\
&& \dgL^{Ze}(\muew)+\dgL^{Z\nu}(\muew)-\dgR^{Ze}(\muew) = -\frac{1}{16\pi^2}\ln\frac{\Lambda}{\muew}\cdot\frac{\sw^2}{\cw^2}(-g^2W+g'^2Y),\qquad \\
&& \dgL^{Zu}(\muew)+\dgL^{Zd}(\muew)-\dgR^{Zu}(\muew)-\dgR^{Zd}(\muew)\CR
&&\qquad = -\frac{1}{16\pi^2}\ln\frac{\Lambda}{\muew}\cdot\frac{1}{3}\frac{\sw^2}{\cw^2} (g^2W-g'^2Y),
\eeqan
\eseq{nu-ew}
where $W, Y, Z$ are the well-defined oblique parameters at the new physics scale (where the theory is universal). Eq.~\eqref{nu-ew} shows that the universal relations \eqref{ur-ew} that hold at LO in universal theories are violated. But unlike generic nonuniversal theories, they are violated in a universal (rather than arbitrary) way. Despite the RG-induced nonuniversal effects, the theory and its phenomenology is still completely characterized by the 16 independent universal parameters at $\mu=\Lambda$ (14 in the limit $y_f\to0$), and no further parameters are needed unlike in generic nonuniversal theories.

As far as observables, or Higgs basis couplings at $\mu=\muew$, are concerned, our discussion in section~\ref{sec:nu} indicates that 
it is not possible to absorb all the LL terms into the running of the oblique parameters that contribute at LO, if $W, Y, Z$ are nonzero at the new physics scale. However, from this perspective, it is convenient to still define $\hat S(\muew)$, $\hat T(\muew)$, $W(\muew)$, $Y(\muew)$ to be their values in the universal limit as in \eqref{STWYewdef}, with $\dot{\hat S}, \dot{\hat T}, \dot W, \dot Y$ given by \eqref{STWYrun}, even beyond this limit when $W(\Lambda), Y(\Lambda), Z(\Lambda)$ are nonzero, so that they can at least absorb a significant part of the LL corrections. The remaining LL corrections are proportional to $W, Y, Z$, and can be taken into account as additional contributions. Following this strategy, we find, for example,
\bseq
\beqa
&&\hspace{-8ex} [\dgL^{Wl}(\muew)]_{ij} = \CR
&&\hspace{-8ex}\quad \frac{\delta_{ij}}{2(\cw^2-\sw^2)} \Bigl\{-2\sw^2\hat S(\muew) +\cw^2\hat T(\muew) -(\cw^2-2\sw^2)W(\muew) +\sw^2Y(\muew) \CR
%&&\qquad +\frac{1}{16\pi^2}\ln\frac{\Lambda}{\muew}\Bigl[\frac{5}{2}(4\cw^2-9\sw^2)g^2W +\Bigl(\frac{11}{6}(3\cw^2-8\sw^2)g'^2 -8\sw^2\lambda\Bigr)Y\Bigr] \Bigr\}, \\
&&\hspace{-8ex}\qquad +\frac{1}{16\pi^2}\ln\frac{\Lambda}{\muew}\Bigl[\Bigl(10g^2-\frac{45}{2}g'^2\Bigr)\cw^2W +\Bigl(\frac{11}{2}g^2-\frac{44}{3}g'^2 -12\lambda\Bigr)\sw^2Y\Bigr] \Bigr\}, \\
&&\hspace{-8ex} [\dgL^{Wq}(\muew)]_{ij} = \CR
&&\hspace{-8ex}\quad \frac{\delta_{ij}}{2(\cw^2-\sw^2)} \Bigl\{-2\sw^2\hat S(\muew) +\cw^2\hat T(\muew) -(\cw^2-2\sw^2)W(\muew) +\sw^2Y(\muew) \CR
%&&\qquad +\frac{1}{16\pi^2}\ln\frac{\Lambda}{\muew} \Bigl[\frac{5}{2}(4\cw^2-9\sw^2)g^2W +\frac{5}{54}(53\cw^2-152\sw^2)g'^2Y +\frac{32}{9}(\cw^2-\sw^2)g_s^2Z\Bigr] \Bigr\},\CR
&&\hspace{-8ex}\qquad +\frac{1}{16\pi^2}\ln\frac{\Lambda}{\muew}\CR
&&\hspace{-8ex}\qquad\quad \Bigl[\Bigl(10g^2-\frac{45}{2}g'^2\Bigr)\cw^2W +\Bigl(\frac{265}{54}g^2-\frac{380}{27}g'^2-12\lambda\Bigr)\sw^2Y +\frac{32}{9}(\cw^2-\sw^2)g_s^2Z\Bigr] \Bigr\},
\eeqan
\eseq{dgLWfnu}
as a generalization of \eqref{dgLWfu}, with the SM parameters still renormalized at $\muew$. These equations quantitatively explain the first example in figure~\ref{fig:diag}. They are obtained by applying the RG equations presented in~\cite{Jenkins:2013zja,Alonso:2013hga} to the full expressions \eqref{dgLWfmu}, and later identifying the various Wilson coefficients involved as combinations of universal parameters, and absorbing part of the LL terms into the running of the oblique parameters according to \eqref{STWYrun}, \eqref{STWYewdef}. Alternatively, \eqref{dgLWfnu} can be more easily derived by realizing that the additional terms compared to \eqref{dgLWfu} can be obtained by turning on $W$, $Y$, $Z$ only (i.e.\ adjusting the Wilson coefficients according to table~\ref{tab:up} to make sure they are the only nonzero universal parameters) when following the steps explained above, and keeping the LL terms. We emphasize that $\hat S(\muew)$, $\hat T(\muew)$, $W(\muew)$, $Y(\muew)$ in \eqref{dgLWfnu} do not have an obvious and unambiguous interpretation in terms of vector boson self-energy corrections, but are simply defined for convenience to absorb {\it part} of the LL corrections. Our prescriptions are by no means the only choice for defining them, but are well-motivated since they leads to relatively simple expressions for the observables and Higgs basis couplings at $\mu=\muew$, such as \eqref{dgLWfnu}.

Finally, we lift the restriction $y_f\to0$ (and meanwhile allowing for nonzero $C_{2JW}$, $C_{2JB}$, $C_{2JG}$). The additional effects come from either the 2 additional operators $Q_y$, $Q_{2y}$, or the $y_f$-dependent contributions to the anomalous dimensions calculated in~\cite{Jenkins:2013wua}, or both. Keeping only the leading terms in $y_t$, we find,
\bseq
\beqa
[\dot C_{Hq}^{(3)}]_{ij} &=& \delta_{ij} \frac{3}{2}y_t^2g^2(C_{HJW}-2C_{2JW}) +\delta_{i3}\delta_{j3}y_t^2 \Bigl(-\frac{1}{2}C_{H\square} +\frac{1}{2}g^2C_{HJW} -\frac{1}{4}g'^2C_{HJB} \CR
&&\quad +\frac{1}{2}g^2C_{2JW} -\frac{1}{18}g'^2C_{2JB} -\frac{8}{3}g_s^2C_{2JG}\Bigr), \\
{[}\dot C_{Hl}^{(3)}]_{ij} &=& \delta_{ij} \frac{3}{2}y_t^2g^2(C_{HJW}-2C_{2JW}), \\
{[}\dot C_{Hq}^{(1)}]_{ij} &=& Y_q \delta_{ij} 3y_t^2g'^2(C_{HJB}-2C_{2JB}) +\delta_{i3}\delta_{j3}y_t^2 \Bigl[\frac{1}{2}(C_{H\square}+C_{HD}) -\frac{9}{4}g^2C_{HJW} \CR
&&\quad +\frac{3}{2}g^2C_{2JW} +\frac{1}{18}g'^2C_{2JB} +\frac{8}{3}g_s^2C_{2JG} +\Bigl(y_t^2-\frac{2}{9}g'^2\Bigr) C_{2y}\Bigr], \\
{[}\dot C_{Hl}^{(1)}]_{ij} &=& Y_l \delta_{ij} 3y_t^2g'^2(C_{HJB}-2C_{2JB}), \\
{[}\dot C_{Hu}]_{ij} &=& Y_u \delta_{ij} 3y_t^2g'^2(C_{HJB}-2C_{2JB}) +\delta_{i3}\delta_{j3}y_t^2 \Bigl[-(C_{H\square}+C_{HD}) +\frac{5}{2}g'^2C_{HJB} \CR
&&\quad -\frac{16}{9}g'^2C_{2JB} -\frac{16}{3}g_s^2C_{2JG} -\Bigl(y_t^2+\frac{1}{9}g'^2\Bigr) C_{2y}\Bigr], \\
{[}\dot C_{Hd}]_{ij} &=& Y_d \delta_{ij} 3y_t^2g'^2(C_{HJB}-2C_{2JB}), \\
{[}\dot C_{He}]_{ij} &=& Y_e \delta_{ij} 3y_t^2g'^2(C_{HJB}-2C_{2JB}).
\eeqan
\eseq{CHfruna}
These should be added to \eqref{CHfrun}. Comparing with \eqref{ur-ew-C}, we see that the additional nonuniversal effects are significant only for the third-generation $q$ and $u$, i.e.\ $t_L$, $b_L$ and $t_R$. They can be represented by the following additional breaking of the universal relations, supplementing \eqref{nu-ew},
\bseq
\beqa
&& [\dgL^{Wq}]_{33} -\dgL^{Wl} = -\frac{y_t^2}{16\pi^2}\ln\frac{\Lambda}{\muew}\biggl[-\frac{3}{2}\frac{\sw^2}{\cw^2}\hat S +\frac{1}{4}\hat T -\frac{9}{4}W +\frac{49}{36}\frac{\sw^2}{\cw^2}Y +\frac{16}{3}\frac{g_s^2}{g^2}Z \CR
&&\qquad -\frac{1}{2}(\cw^2+3\sw^2)\Dgzb +\frac{3}{2}\frac{\sw^2}{\cw^2}\Dkapab -\frac{1}{2}\DkapVb\biggr] \label{nuewa1}\\
&& \frac{[\dgR^{Zu}]_{33}}{Y_u} -\frac{\dgR^{Zd}}{Y_d} = -\frac{y_t^2}{16\pi^2}\ln\frac{\Lambda}{\muew} \biggl[-\frac{15}{4}\frac{\sw^2}{\cw^2}\hat S -\frac{15}{8}\hat T -\frac{9}{8}W +\frac{71}{24}\frac{\sw^2}{\cw^2}Y -8\frac{g_s^2}{g^2}Z \CR
&&\qquad -\frac{3}{4}(3\cw^2+5\sw^2)\Dgzb +\frac{15}{4}\frac{\sw^2}{\cw^2}\Dkapab +\frac{3}{4}\DkapVb +\frac{1}{12}(9y_t^2 +g'^2)c_{2y}\biggr], \\
&& {[}\dgL^{Zu}]_{33}+[\dgL^{Zd}]_{33}-[\dgR^{Zu}]_{33}-\dgR^{Zd} = -\frac{y_t^2}{16\pi^2}\ln\frac{\Lambda}{\muew} \biggl[\frac{5}{2}\hat T -\frac{9}{2}W -\frac{11}{18}\frac{\sw^2}{\cw^2}Y +\frac{32}{3}\frac{g_s^2}{g^2}Z \CR
&&\qquad -6\cw^2\Dgzb -\DkapVb -\frac{1}{6}(9y_t^2 -g'^2) c_{2y}\biggr].
\eeqan
\eseq{nuewa}
The other universal relations are not violated up to $y_f^2/y_t^2(f\ne t)$ suppressed terms. Note also that, as indicated above, $[\dgR^{Zd}]_{33}$ is not modified by terms proportional to $y_t^2$, so $[\dgR^{Zd}]_{ij}\propto\delta_{ij}$ still holds approximately.

The universal pieces in \eqref{CHfruna}, on the other hand, can be conveniently attributed to the running of $C_{HJW}$, $C_{HJB}$ in addition to \eqref{CHJVrun},
\bseq
\beqa
\dot C_{HJW} &=& 6y_t^2(C_{HJW}-2C_{2JW}), \\
\dot C_{HJB} &=& 6y_t^2(C_{HJB}-2C_{2JB}).
\eeqan
\eseqn
Note that the one-loop beta functions of $g$, $g'$ do not depend on $y_t$. Regarding the 4-fermion interactions related to the $W, Y, Z$ parameters, the additional contributions to the anomalous dimensions are significant only for the third-generation quarks $t_L$, $b_L$, $t_R$, and there is no universal part to be added to $\dot C_{2JW}$, $\dot C_{2JB}$, $\dot C_{2JG}$. Further, the running of $C_{HWB}$, $C_{HD}$ in \eqref{Cbosrun} should be supplemented by the following additional terms, taken from~\cite{Jenkins:2013wua},
\bseq
\beqa
\dot C_{HWB} &=& 6y_t^2 C_{HWB}, \\
%\dot C_{HD} &=& 12(y_t^2+\lambda)C_{HD} -6y_t^2g'^2C_{HJB},
\dot C_{HD} &=& 6y_t^2 (2C_{HD} -g'^2C_{HJB}),
\eeqan
\eseqn

The discussion above implies that up to nonuniversal effects that are important for the third-generation quarks $t_L$, $b_L$, $t_R$ only, the $y_f$-dependent RG effects in the electroweak sector are universal and can be conveniently attributed to the running of the oblique parameters. Referring to table~\ref{tab:up} for the translation between the universal parameters and the Warsaw basis Wilson coefficients, we find
%\bseq
%\beqa
%\dot{\hat S} &=& 6y_t^2\hat S +4\lambda\Dkapab, \\
%\dot{\hat T} &=& -24\lambda\frac{\sw^2}{\cw^2}\hat S +12(y_t^2+\lambda)\hat T +12\lambda\frac{\sw^2}{\cw^2}Y -24\lambda\sw^2\Dgzb +24\lambda\frac{\sw^2}{\cw^2}\Dkapab, \\
%\eeqan
%\eseq{STWYruna}
\beq
\dot{\hat S} = 6y_t^2\hat S,\quad \dot{\hat T} = 12y_t^2\hat T,\quad \dot W = \dot Y = 0.
\eeq{STWYruna}
These equations are to be added to \eqref{STWYrun}. Similarly, we still have $\dot Z=0$. We remark in passing that \eqref{STWYrun} and \eqref{STWYruna} can also be derived from the results in~\cite{Elias-Miro:2013eta}, where the submatrix of $\gamma_{ij}$ involving the bosonic operators in the EGGM basis is calculated. Referring to~\cite{Wells:2015uba} for the expressions of the universal parameters in this basis, we have explicitly checked that the results are the same as we presented above.

Defined in this way, the oblique parameters that appear in the LO expressions of electroweak observables, when renormalized at $\mu=\muew$, absorb all the $\O(\frac{y_t^2}{16\pi^2}\ln\frac{\Lambda}{\muew})$ corrections, except for observables involving the $Z$ boson couplings to $t_L$, $b_L$ and $t_R$. Among them, only the $Zb_L\bar b_L$ coupling is directly probed by precision $Z$-pole data, for which we obtain (suppressing the gauge-coupling-dependent LL corrections proportional to $W$, $Y$, $Z$ discussed in the previous subsection),
\beqa
&&[\dgL^{Zd}(\muew)]_{33} \CR
&&= \frac{1}{12(\cw^2-\sw^2)} \Bigl[4\sw^2\hat S(\muew) -(3 -4\sw^2)\hat T(\muew) +(3-8\sw^2)W(\muew) -\frac{\sw^2}{\cw^2}Y(\muew)\Bigr] \CR
&&\quad +\frac{y_t^2}{32\pi^2}\ln\frac{\Lambda}{\muew}\Bigl[\frac{\sw^2}{\cw^2}(\hat S-\Dkapab) -\hat T +3W +(7-6\sw^2)\Dgzb +\Bigl(y_t^2-\frac{2}{9}g'^2\Bigr)c_{2y}\Bigr].
\eeqa{dgLZdnu}
The physical picture of this effect was already discussed in the second example in figure~\ref{fig:diag}.

%%%
\subsection{Implications for the oblique parameters fit}

So far, we have found that while universal theories at the new physics scale do not in general remain universal after RG evolution down to the electroweak scale, precision observables in the electroweak sector allow for a separation of universal and nonuniversal effects induced by RG evolution. With our prescriptions for the separation, the universal effects are conveniently attributed to the running of the oblique parameters, given by the sum of \eqref{STWYrun} and \eqref{STWYruna}. This serves as a definition of the oblique parameters at the electroweak scale; see \eqref{STWYewdef}. Corrections to the electroweak observables not involving the third-generation quarks $t_L$, $b_L$, $t_R$ can be represented, to LL and leading $y_t$ accuracy, by the LO expressions with $\hat S$, $\hat T$, $W$, $Y$ renormalized at $\muew$, plus additional (nonuniversal) terms proportional to $\frac{1}{16\pi^2}\ln\frac{\Lambda}{\muew}\cdot\{W,\, Y,\, Z\}$; see e.g.\ \eqref{dgLWfnu}. For the electroweak observables involving $t_L$, $b_L$, or $t_R$, on the other hand, additional terms of order $\frac{y_t^2}{16\pi^2}\ln\frac{\Lambda}{\muew}$ should be added, which also involve some less-constrained nonoblique universal parameters; see e.g.\ \eqref{dgLZdnu}.

If these additional LL terms were absent (or negligible), the conventional oblique parameters fit, where theory predictions of observables incorporating LO contributions from the oblique parameters are confronted with precision electroweak data, would be a consistent procedure to derive constraints on universal theories. Bounds on the oblique parameters obtained in this way could be interpreted as bounds on $\hat S(\muew)$, $\hat T(\muew)$, $W(\muew)$, $Y(\muew)$ defined in \eqref{STWYewdef}. The latter could then be mapped onto constraints on the universal parameters at the new physics scale $\Lambda$, following the sum of \eqref{STWYrun} and \eqref{STWYruna}. 

In reality, however, the additional LL terms due to RG-induced nonuniversal effects, which involve some less-constrained universal parameters, may not be negligible compared with LO contributions from $\hat S$, $\hat T$, $W$, $Y$, as well as experimental and SM theoretical uncertainties. If this is the case, one should go beyond LO for a consistent fit of universal theories to precision electroweak data. But as far as universal theories are concerned, the underlying number of free parameters is still much smaller than that in the full SMEFT. At the LL order, only a few additional parameters, defined by linear combinations of the universal parameters at $\Lambda$, are sufficient. While a full-fledged global analysis is beyond the scope of the present paper, we will illustrate this point with an example in the next subsection.

\subsection{Example: $\Rl$ and $\Rb$ in universal theories}

We consider the two observables $\Rl$ and $\Rb$ introduced at the end of section~\ref{sec:nu}, and see how their SMEFT predictions are affected by the additional nonuniversal LL terms. Similar to the examples shown in the previous subsections, namely \eqref{dgLWfnu} and \eqref{dgLZdnu}, the Higgs basis couplings renormalized at $\muew$ that appear in \eqref{dbNPRlRb} can be worked out. Eq.~\eqref{dbNPRlRb} then becomes, numerically,
\bseq
\beqa
\dbNP\Rl &=& -0.36\bigl[\De_3(\muew)-\cw^2\De_1(\muew)\bigr] \CR
&&\quad +\frac{\ln(\Lambda/\muew)}{3}(0.13Z -0.053\Dgzb +0.0028\Dkapab -0.0091c_{2y}), \\
\dbNP\Rb &=& 0.079 \bigl[\De_3(\muew)-\cw^2\De_1(\muew)\bigr] \CR
&&\quad +\frac{\ln(\Lambda/\muew)}{3}(-0.19\Dgzb +0.010\Dkapab -0.032c_{2y}),
\eeqan
\eseq{RlRbnum}
where
\beq
\De_3(\muew)-\cw^2\De_1(\muew) = \hat S(\muew) -0.77\hat T(\muew) -0.23W(\muew) -0.77Y(\muew)
\eeq{oblcomb}
is a common oblique parameters combination entering the two observables at LO, expressed in terms of the $\De$ parameters defined in \eqref{Dedef}.\footnote{With only observables involving ratios of $Zf\bar f$ couplings such as $\Rl$ and $\Rb$, one cannot break this degeneracy, because $g_i^{Zf}+\delta g_i^{Zf} = (1+\frac{\De_1}{2})g_i^{Zf} -Q_f\frac{\sw^2}{\cw^2-\sw^2}(\De_3-\cw^2\De_1)$, for both $i=L,R$. When $\De_3-\cw^2\De_1=0$, all $Zf\bar f$ couplings are rescaled by a common factor, and ratios of couplings are unchanged. This flat direction can be lifted by considering other observables such as the $Z$ boson total width.} We have neglected the additional LL terms proportional to $\hat S$, $\hat T$, $W$, $Y$, since these parameters already appear in the LO expressions. The numerical impact of these neglected terms is to correct the coefficients of $\De_3(\muew)-\cw^2\De_1(\muew)$ by order $\frac{1}{16\pi^2}\ln\frac{\Lambda}{\muew}$ numbers, and is expected to be less important than the invasion of additional, possibly less-constrained parameters $Z$, $\Dgzb$, $\Dkapab$, $c_{2y}$ through RG evolution from $\Lambda$ to $\muew$.

\begin{figure}[tbp]
\centering
\includegraphics[width=.95\textwidth]{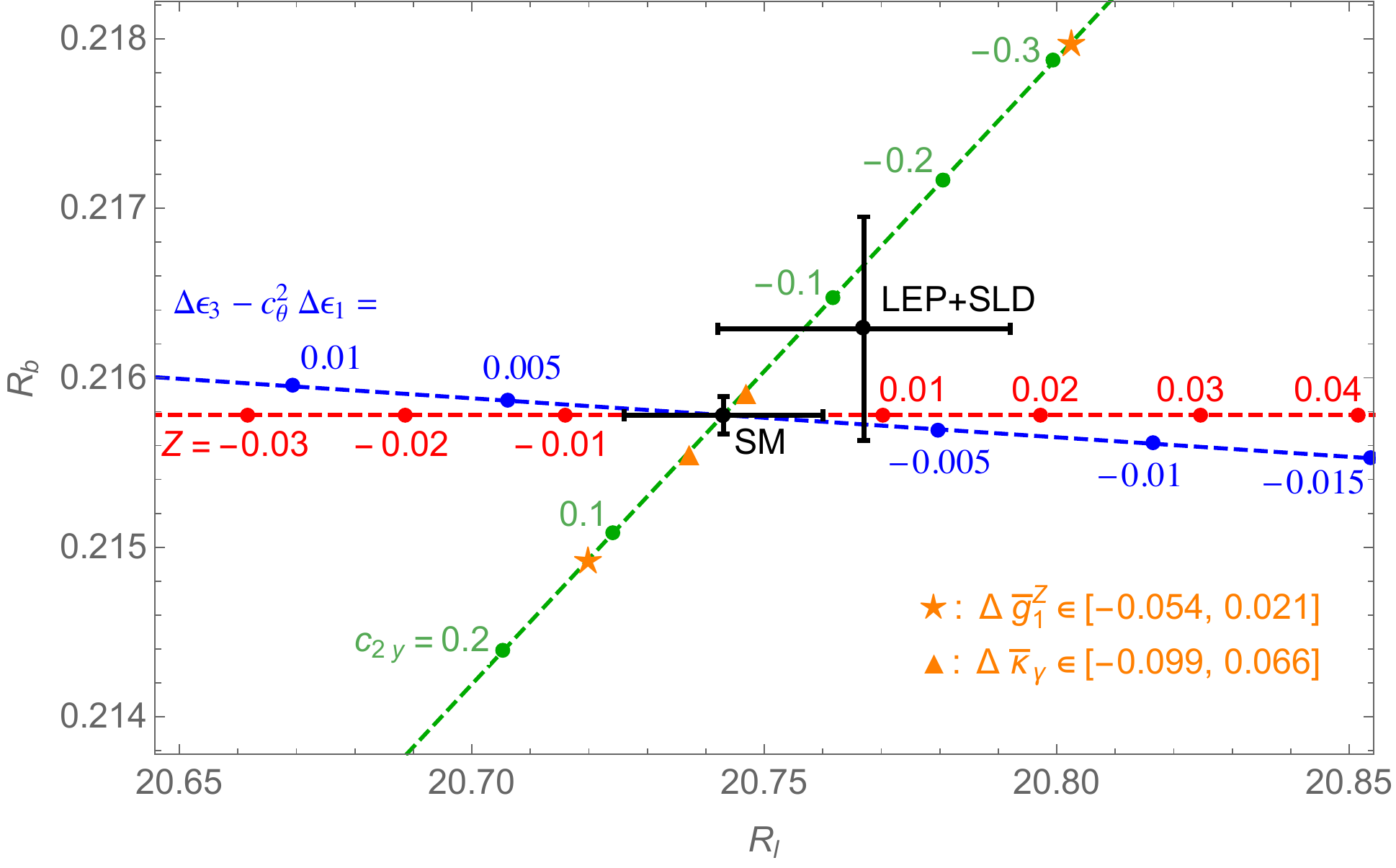}\hspace{.05\textwidth}
\caption{\label{fig:RlRb} Theory predictions for $\Rl=\Gamhad/\Gamma(Z\to\ell^+\ell^-)$ and $\Rb=\Gamma(Z\to b\bar b)/\Gamhad$ are shifted away from the SM point along the dashed lines, when the universal parameters appearing in \eqref{RlRbnum} take the values labeled beside the dots. The anomalous TGC parameters $\Dgzb$, $\Dkapab$ lead to shifts along the same direction as $c_{2y}$ (green dashed line), with the orange stars and triangles indicating the maximum shifts allowed by the LEP2 TGC constraints (95\%~C.L.) from single-parameter fits (shown in the bottom-right corner)~\cite{Schael:2013ita}. $\ln\frac{\Lambda}{\muew}=3$ is assumed, as motivated by TeV-scale new physics. Agreement between the SM predictions as fitted by the Gfitter group~\cite{Baak:2014ora} and the combined measurements by the LEP and SLD collaborations~\cite{ALEPH:2005ab} naively constrains the oblique parameters combination $\De_3-\cw^2\De_1$ (blue) defined in \eqref{oblcomb} at the $10^{-3}$ level. But even when the oblique parameters are interpreted as renormalized at $\muew$ following our prescriptions, the neglected LL terms in such a LO oblique parameters analysis can actually be significant. %, when reasonable values of order $\frac{v^2}{\Lambda^2}$ are assumed for $Z$ (red) and $c_{2y}$ (green), and the quoted constraints on $\Dgzb$, $\Dkapab$ are satisfied. 
The challenge illustrated by this example requires extending the $(\hat S, \hat T, W, Y)$ parametrization to include additional parameters in a consistent global fit of universal theories beyond LO.}
\end{figure}

The various terms in \eqref{RlRbnum} shift the theory predictions for $\Rl$ and $\Rb$ in different directions in the $\Rl$-$\Rb$ plane. This is shown by the dashed lines in figure~\ref{fig:RlRb}, assuming $\ln\frac{\Lambda}{\muew}=3$ as expected from $\Lambda\sim\O(\text{TeV})$. The new physics corrections can be compared with the SM predictions from the Gfitter fit~\cite{Baak:2014ora},
\beq
\Rl = 20.743\pm0.017,\quad \Rb = 0.21578\pm0.00011,\quad\text{(SM)}\qquad\quad
\eeqn
which is based on the $Z$-pole measurements from the LEP and SLD collaborations~\cite{ALEPH:2005ab},
\beq
\Rl = 20.767\pm0.025,\quad \Rb = 0.21629\pm0.00066.\quad\text{(LEP+SLD)}
\eeqn

As we can see from figure~\ref{fig:RlRb}, a LO oblique parameters fit would naively constrain the linear combination $\De_3-\cw^2\De_1$ (blue), properly renormalized at $\muew$, to be $\O(10^{-3})$. However, reasonable values of other universal parameters, namely $\O(\frac{v^2}{\Lambda^2})$, which enter the LL corrections, can significantly change the picture. In particular, values of $\O(10^{-2})$ and $\O(10^{-1})$ for the $Z$ (red) and $c_{2y}$ (green) parameters, respectively, which may be generated by heavy QCD-charged states and scalar states, lead to corrections larger than the experimental and SM theoretical uncertainties. It would be interesting to compare these numbers with direct constraints on the parameters $Z$ (see e.g.~\cite{Domenech:2012ai}) and $c_{2y}$, and obtain a fuller understanding of allowed parameter ranges through a global SMEFT analysis. The anomalous TGC parameters $\Dgzb$ and $\Dkapab$ shift the theory predictions along the same direction as $c_{2y}$, since all three parameters contribute via $[\dgL^{Zd}]_{33}$ only. They are directly constrained by measurements at LEP2, and more recently also at the LHC.\footnote{Though experimental constraints are on $\dgz$, $\dkapa$ defined with respect to the physical particles, the difference between $\dgz$ and $\Dgzb$, which involve $\De_{1,2,3}$ (see table~\ref{tab:hb}), is not relevant, since when interpreted in universal theories, oblique corrections are always assumed to vanish in experimental TGC analyses. See~\cite{Wells:2015uba} for more discussion.} The green line segment between the orange stars (triangles) represents the 95\%~C.L.\ interval allowed by the combined LEP2 constraint on $\Dgzb$ ($\Dkapab$) taken from the LEP electroweak working group final report~\cite{Schael:2013ita}. 
These constraints are derived allowing one anomalous TGC parameter to be nonzero at a time, and are shown here for illustration purpose only. We see that values of $\Dgzb$ as allowed by the above constraint can contribute significant corrections to $\Rl$ and $\Rb$.

Our example shows that the RG-induced nonuniversal effects that are usually neglected can indeed challenge the interpretation and usefulness of the LO oblique parameters analysis. In practice, this means that for a consistent global fit of universal theories to precision electroweak data, one should go beyond the conventional approach with the $(\hat S, \hat T, W, Y)$ parametrization. An extension to LL order should at least involve two additional parameters,
\beq
\tilde Z \equiv Z\,\ln\frac{\Lambda}{\muew},\quad
\delta\tilde g_L^{Zb} \equiv \Bigl[(7-6\sw^2)\Dgzb -\frac{\sw^2}{\cw^2}\Dkapab +\Bigl(y_t^2-\frac{2}{9}g'^2\Bigr)c_{2y}\Bigr]\ln\frac{\Lambda}{\muew},
\eeq{addparam}
where $\delta\tilde g_L^{Zb}$ is proportional to the linear combination of the less-constrained universal parameters appearing in the LL term in \eqref{dgLZdnu}. For the two observables $\Rl$ and $\Rb$ discussed in this subsection, $\tilde Z$ and $\delta\tilde g_L^{Zb}$ capture shifts in the directions of the red and green dashed lines in figure~\ref{fig:RlRb}, respectively. Further extending the analysis to include NLO finite corrections may introduce more parameters, but the total number of free parameters is no more than 16, the number of universal parameters defined at $\Lambda$.\footnote{A further challenge can potentially arise at this order, if constraints on the universal parameters are to be interpreted in specific UV models. %While our EFT definition of universal theories makes no reference to UV completions, we note that UV theories that are usually regarded as universal may not match exactly onto the universal theories EFT, if the matching is performed beyond tree level. 
Since a NLO calculation of observables requires one-loop matching~\cite{Henning:2014wua,Drozd:2015kva,Chiang:2015ura,Huo:2015exa,Huo:2015nka,Drozd:2015rsp} of the Wilson coefficients contributing at LO, we need to assume that the UV model does not generate operators beyond those in \eqref{Lu} even at one-loop matching. This assumption is implicit in our EFT definition of universal theories, but may not be satisfied by all UV theories that would otherwise be regarded as universal.} Extended in this way, the oblique parameters analysis can be consistent and useful, and yet simpler than the full SMEFT if one is interested only in universal theories (see~\cite{Berthier:2015oma,Berthier:2015gja} for discussions on consistent analyses of the full SMEFT).

%%%
\section{RG effects in the Yukawa sector}
\label{sec:yukawa}

We next turn to the Yukawa sector, and show how the universal relation \eqref{ur-yuk} can be violated by RG evolution. The observation that RG evolution in universal theories can induce nonuniversal rescaling of all SM fermion Yukawa couplings was previously made in~\cite{Elias-Miro:2013mua}, based on partial results on the anomalous dimensions $\gamma_{ij}$ for one fermion generation, and assuming a limited set of nonzero Wilson coefficients. Our analysis in this section takes into account the full $\gamma_{ij}$ that became available after~\cite{Elias-Miro:2013mua}, and all the parameters characterizing universal theories classified in~\cite{Wells:2015uba}.

The dimension-6 operators relevant for Yukawa coupling corrections are those in the $\psi^2H^3$ class. At LO, their Wilson coefficients are related in universal theories as follows,
\beq
\bigl[ \{C_{uH}, C_{dH}, C_{eH}\} \bigr]_{ij} = \bigl[\{y_u, \VCKM y_d, y_e\}\bigr]_{ij} C_y.
\eeqn
The running of these Wilson coefficients is in general complicated by the nontrivial flavor structure in the quark sector. For example, $[\dot C_{dH}]_{ij}$ contains terms proportional to $[y_u y_u^\dagger \VCKM y_d]_{ij}$, which, unlike $[\VCKM y_d]_{ij}$ that $[C_{dH}]_{ij}$ is proportional to at LO, cannot be diagonalized by applying $\VCKM^\dagger$ on the left. Thus, a redefinition of the CKM matrix is needed after RG evolution. However, the third-generation quarks are hardly affected by this complication, since we can approximate $\VCKM$ by a block-diagonal matrix,
\beq
\VCKM \simeq \left(
\begin{matrix}
1 & \lambda_W & \,\,\,\,0\,\, \\
-\lambda_W & 1 & \,\,\,\,0\,\, \\
0 & 0 & \,\,\,\,1\,\, 
\end{matrix}
\right),
\eeq{VCKMapprox}
where a subscript ``W'' has been added to the Wolfenstein parameter $\lambda_W\simeq0.23$ to avoid confusion with the Higgs self-coupling $\lambda$. With $\O(\lambda_W^2)$ terms neglected, RG evolution in universal theories does not mix the third-generation quarks with the first- and second-generation ones. We will focus on the experimentally most accessible third-generation Yukawa coupling corrections in the following, adopting the approximation \eqref{VCKMapprox} and neglecting terms suppressed by $y_f^2/y_t^2\,(f\ne t)$. Using the results in~\cite{Jenkins:2013zja,Jenkins:2013wua,Alonso:2013hga} and table~\ref{tab:Qf}, we find
\bseq
\beqa
[\dot C_{uH}]_{33} &=& y_t \biggl[ \Bigl(\frac{51}{2}y_t^2+24\lambda-8g_s^2-\frac{27}{4}g^2-\frac{35}{12}g'^2\Bigr)C_y -12y_t^2(y_t^2-\lambda)C_{2y} \CR
&&\quad -(3y_t^2-3\lambda-4g^2+g'^2)g^2C_{HJW} +\Bigl(\frac{1}{2}y_t^2+\lambda-g^2+\frac{2}{3}g'^2\Bigr)g'^2C_{HJB} \CR
&&\quad +\frac{16}{9}(y_t^2-\lambda) g'^2C_{2JB} +\frac{64}{3}(y_t^2-\lambda) g_s^2C_{2JG} \CR
&&\quad -\Bigl(6y_t^2+4\lambda-\frac{10}{3}g^2\Bigr)C_{H\square} +\Bigl(y_t^2+2\lambda-\frac{3}{2}g^2+\frac{3}{2}g'^2\Bigr)C_{HD} \CR
&&\quad -gg'C_{HWB} +32g_s^2C_{HG} +9g^2C_{HW} +\frac{17}{3}g'^2C_{HB} \biggr], \\
{[}\dot C_{dH}]_{33} &=& y_b \biggl[\Bigl(\frac{21}{2}y_t^2+24\lambda-8g_s^2-\frac{27}{4}g^2-\frac{23}{12}g'^2\Bigr)C_y -14y_t^2(y_t^2-\lambda)C_{2y} \CR
&&\quad -\Bigl(\frac{3}{2}y_t^2-3\lambda-4g^2+\frac{1}{2}g'^2\Bigr)g^2C_{HJW} +\Bigl(\lambda-\frac{1}{2}g^2-\frac{1}{3}g'^2\Bigr)g'^2C_{HJB} \CR
&&\quad  +\frac{8}{9}\lambda g'^2C_{2JB} -\frac{64}{3}\lambda g_s^2C_{2JG} -\Bigl(4\lambda-\frac{10}{3}g^2\Bigr)C_{H\square} +\Bigl(2\lambda-\frac{3}{2}g^2+\frac{3}{2}g'^2\Bigr)C_{HD} \CR
&&\quad +gg'C_{HWB} +32g_s^2C_{HG} +9g^2C_{HW} +\frac{5}{3}g'^2C_{HB} \biggr], \\
{[}\dot C_{eH}]_{33} &=& y_\tau \biggl[\Bigl(15y_t^2+24\lambda-\frac{27}{4}g^2-\frac{21}{4}g'^2\Bigr)C_y -12y_t^2(y_t^2-\lambda)C_{2y} \CR
&&\quad -\Bigl(3y_t^2-3\lambda-4g^2+\frac{3}{2}g'^2\Bigr)g^2C_{HJW} +\Bigl(\lambda-\frac{3}{2}g^2+3g'^2\Bigr)g'^2C_{HJB} \CR
&&\quad -8\lambda g'^2C_{2JB} -\Bigl(4\lambda-\frac{10}{3}g^2\Bigr)C_{H\square} +\Bigl(2\lambda-\frac{3}{2}g^2+\frac{3}{2}g'^2\Bigr)C_{HD} \CR
&&\quad -3gg'C_{HWB} +9g^2C_{HW} +15g'^2C_{HB} \biggr].
\eeqan
\eseq{CfHrun}
While there are overlapping terms in these equations, there is no obvious well-motivated way to make the separation between universal vs.\ nonuniversal effects. We thus refrain from defining the running of $\DkapFb$ as we did for the oblique parameters in the previous section, but simply present the violation of the universal relation \eqref{ur-yuk} at the electroweak scale. To do so, we note that, in our notation,
%\beq
%\frac{[C_{uH}]_{33}}{y_t} -\frac{[C_{dH}]_{33}}{y_b} = -(\delta y_t -\delta y_b),\quad
%\frac{[C_{dH}]_{33}}{y_b} -\frac{[C_{eH}]_{33}}{y_\tau} = -(\delta y_b -\delta y_\tau),
%\eeqn
\beq
\delta y_t -\delta y_b = -\biggl(\frac{[C_{uH}]_{33}}{y_t} -\frac{[C_{dH}]_{33}}{y_b}\biggr),\quad
\delta y_b -\delta y_\tau = -\biggl(\frac{[C_{dH}]_{33}}{y_b} -\frac{[C_{eH}]_{33}}{y_\tau}\biggr),
\eeqn
where $\delta y_t$, $\delta y_b$, $\delta y_\tau$ represent $[\delta y_u]_{33}$, $[\delta y_d]_{33}$, $[\delta y_e]_{33}$, respectively; see \eqref{dyf}. Combining \eqref{CHfrun} and the one-loop running of the SM Yukawa couplings,
\bseq
\beqa
\dot y_t &=& y_t \Bigl(\frac{9}{2}y_t^2 -8g_s^2 -\frac{9}{4}g^2 -\frac{17}{12}g'^2\Bigr), \\
\dot y_b &=& y_b \Bigl(\frac{3}{2}y_t^2 -8g_s^2 -\frac{9}{4}g^2 -\frac{5}{12}g'^2\Bigr), \\
\dot y_\tau &=& y_\tau \Bigl(3y_t^2 -\frac{9}{4}g^2 -\frac{15}{4}g'^2\Bigr),
\eeqan
\eseqn
we obtain
\bseq
\beqa
&& \delta y_t(\muew) -\delta y_b(\muew) = -\frac{1}{16\pi^2}\ln\frac{\Lambda}{\muew} (\dot{\delta y_t} -\dot{\delta y_b}) \CR
&&\quad = \frac{1}{16\pi^2}\ln\frac{\Lambda}{\muew} \Bigl[ -6y_t^2(2\DkapFb -\DkapVb) +4g'^2\sw^2\Dgzb -2g'^2\frac{\sw^2}{\cw^2}\Dkapab \CR
&&\qquad -2(g^2-2g'^2)\frac{\sw^2}{\cw^2}\hat S +y_t^2\hat T +(3y_t^2+2g'^2)W -\Bigl(\frac{41}{9}y_t^2-\frac{16}{3}\lambda-2g^2+4g'^2\Bigr)\frac{\sw^2}{\cw^2}Y \CR
&&\qquad -\frac{128}{3}y_t^2\frac{g_s^2}{g^2}Z +2(y_t^2-\lambda)y_t^2c_{2y} +g'^2(e^2f_{\gamma\gamma} -g'^2f_{z\gamma}) \Bigr] \CR
&&\quad \simeq \frac{\ln(\Lambda/\muew)}{3} (-0.23\DkapFb +0.11\DkapVb +0.0022\Dgzb -0.0014\Dkapab -0.0019\hat S +0.019\hat T \CR
&&\qquad +0.061W -0.020Y -2.8Z +0.032c_{2y} +0.00023f_{\gamma\gamma} -0.00031f_{z\gamma}), \\
&& \delta y_b(\muew) -\delta y_\tau(\muew) = -\frac{1}{16\pi^2}\ln\frac{\Lambda}{\muew} (\dot{\delta y_b} -\dot{\delta y_\tau}) \CR
&&\quad = \frac{1}{16\pi^2}\ln\frac{\Lambda}{\muew} \Bigl[3y_t^2(\DkapFb -\DkapVb) -\frac{40}{3}g'^2\sw^2\Dgzb +\frac{20}{3}g'^2\frac{\sw^2}{\cw^2}\Dkapab \CR
&&\qquad +4\Bigl(g^2-\frac{10}{3}g'^2\Bigr)\frac{\sw^2}{\cw^2}\hat S -\Bigl(3y_t^2 +4g'^2\Bigr)W -4\Bigl(\frac{40}{9}\lambda+g^2-\frac{10}{3}g'^2\Bigr)\frac{\sw^2}{\cw^2}Y \CR
&&\qquad +\frac{128}{3}\lambda\frac{g_s^2}{g^2}Z -2(y_t^2-\lambda)y_t^2c_{2y} +8g_s^4 f_{gg} -\frac{10}{3}g'^2(e^2f_{\gamma\gamma}-g'^2f_{z\gamma}) \Bigr] \CR
&&\quad \simeq \frac{\ln(\Lambda/\muew)}{3} (0.056\DkapFb -0.056\DkapVb -0.0074\Dgzb +0.0048\Dkapab -0.000014\hat S \CR
&&\qquad -0.066W -0.013Y +0.37Z -0.032c_{2y} +0.34f_{gg} -0.00078f_{\gamma\gamma} +0.0010f_{z\gamma}).
\eeqan
\eseq{nu-yuk}
The terms in these equations involving the oblique parameters correspond to the effect illustrated by the third example in figure~\ref{fig:diag}. 

The numerical results in \eqref{nu-yuk} show that significant deviations from the universal relation \eqref{ur-yuk} are possible. For example, in the simplest scenario where $\DkapFb$ is the only nonnegligible universal parameter at the new physics scale $\Lambda$, we have $\delta y_t(\Lambda) = \delta y_b(\Lambda) = \delta y_\tau(\Lambda) = \DkapFb$, but $\delta y_t(\muew) \simeq 0.77\delta y_b(\muew)$, $\delta y_b(\muew) \simeq 1.056\delta y_\tau(\muew)$ after RG evolution, if $\ln\frac{\Lambda}{\muew}\simeq3$. %\com{These numbers still differ from~\cite{Elias-Miro:2013mua}. The difference seems to originate from $\dot C_{dH}$.} 
Further deviations can be induced by other universal parameters, such as $\DkapVb$, $Z$, $c_{2y}$, $f_{gg}$, if they are generated at $\Lambda$. Therefore, the sometimes adopted simplified approach to precision Higgs fit where a common rescaling factor is assumed for all the SM fermion Yukawa couplings does not find its justification in universal theories. This assumption applies to the the effective $hff$ couplings at $\mu\sim m_h\sim\muew$, and appears fine-tuned in light of the RG-induced nonuniversal effects illustrated above. Thus, even for universal theories, it is desirable to keep these parameters separate when fitting them to data.

%%%
\section{Conclusions}
\label{sec:conclusions}

The usefulness of simplified frameworks for precision analyses lies in the fact that they are much more tractable than the full SMEFT with a vast parameter space, and yet capture broad classes of BSM scenarios. The oblique parameters framework, which characterizes effects of universal theories on precision electroweak observables, has been widely-used for more than two decades now, and finds its justification at LO in the modern SMEFT approach with a consistent description of universal theories in the SMEFT~\cite{Wells:2015uba}. In many cases, however, it is desirable to go beyond LO in the new physics effects, and simplified frameworks should be properly extended to incorporate RG evolution.

In this paper, we have performed a RG analysis of universal theories in the SMEFT framework. The key observation is that under RG evolution, universal theories at the new physics scale $\Lambda$, which reside in a 16-dimensional subspace of the full SMEFT parameter space, can flow out of this subspace, and become nonuniversal at the electroweak scale $\muew$ where their effects on precision observables are measured. But the departure from universal theories at $\muew$ is not arbitrary, as the theory is still usefully described by the 16 universal parameters defined at $\Lambda$. The main consequences of this observation are the following.\begin{itemize}
\item The universal pattern of deviations from SM predictions seen at LO in the universal theories EFT is distorted after RG evolution from $\Lambda$ to $\muew$. The RG-induced nonuniversal effects lead to well-defined departures (dictated by the 16 universal parameters at $\Lambda$) 
from the LO universal relations \eqref{ur} among some generically independent Higgs basis couplings (in the sense explained at the end of section~\ref{sec:nu}); see \eqref{nu-ew}, \eqref{nuewa}, \eqref{nu-yuk}.
\item Since there is in general no unique procedure to define the oblique parameters (and more generally universal parameters) for nonuniversal theories, additional prescriptions are needed for $\hat S(\muew)$, $\hat T(\muew)$, etc.\ to be meaningful. Our prescriptions are shown in \eqref{STWYewdef}, where the running of the oblique parameters is given by the sum of \eqref{STWYrun} and \eqref{STWYruna}.
\item With our prescriptions, LO expressions for the new physics corrections to electroweak observables $\dbNP\Obs$ can be used with $\hat S$, $\hat T$, $W$, $Y$ renormalized at $\muew$, supplemented by additional LL terms that cannot be absorbed into the running of the oblique parameters. An example calculation of two well-measured observables $\Rl$ and $\Rb$ shows that the additional LL terms can be numerically important; see \eqref{RlRbnum} and figure~\ref{fig:RlRb}. This implies that, even for universal theories, a consistent precision electroweak fit should go beyond the $\{\hat S, \hat T, W, Y\}$ parametrization. But unlike generic nonuniversal theories, the additional parameters to be incorporated are a small number of linear combinations of other universal parameters invading through RG evolution from $\Lambda$ to $\muew$; see \eqref{addparam}.
\item The Yukawa couplings of all SM fermions are in general not modified in the same way even in universal theories. In particular, \eqref{nu-yuk} shows the potentially sizable RG-induced deviations from a universal rescaling for the top, bottom and tau Yukawa couplings (as parameters in the Higgs basis framework). Thus, fitting a common Yukawa coupling rescaling factor to Higgs data as based on LO intuitions from universal theories is of limited use.
\end{itemize}

Two additional aspects of RG-induced nonuniversal effects not discussed in this paper are the generation of the dipole-type couplings $d_{Vf}$ (which vanish at LO in universal theories; see table~\ref{tab:hb}), and a nonuniversal pattern of 4-fermion interactions. They correspond to violations of the two other features of universal theories at LO listed in section 4.2 of~~\cite{Wells:2015uba} that are not captured by the universal relations \eqref{ur}.\footnote{There it is also mentioned that $\dgR^{Wq}=0$ at LO in universal theories; see table~\ref{tab:hb} of the present paper. A nonzero $\dgR^{Wq}$ is generated by RG evolution at $\O(y_uy_d)$.} Following the discussion in~\cite{Alonso:2013hga}, we see the former affects the muon anomalous magnetic moment, but not $\mu\to e\gamma$ or electric dipole moments, if the theory is universal (and CP-conserving) at $\Lambda$. The latter aspect may have an impact on precision analyses of LEP2 data in the oblique parameters framework, and can also be relevant for future precision measurements on a higher-energy $e^+e^-$ collider where also the top quark can be pair-produced. In any case, to make maximal use of existing and upcoming precision data for indirect searches of physics beyond the SM, simplified parameterizations of new physics effects, as motivated by specific classes of BSM scenarios like universal theories, should be consistently cast in the SMEFT framework (if the absence of new light states is assumed), and checked for robustness against RG evolution.

%%%
\acknowledgments
We thank S.~Martin and M.~Trott for communications. This work is supported in part by the the U.S.\ Department of Energy under grant DE-SC0007859.

%\paragraph{Note added.} 

%%%
\appendix

%%%
\section{Notation for the SMEFT}
\label{app:notation}

In our notation, the SM Lagrangian reads
\beqa
\L_\SM &=& -\frac{1}{4}G^A_{\mu\nu}G^{A\mu\nu} -\frac{1}{4}W^a_{\mu\nu}W^{a\mu\nu} -\frac{1}{4}B_{\mu\nu}B^{\mu\nu} +|D_\mu H|^2 +\lambda v^2|H|^2 -\lambda|H|^4 \CR
&& +\sum_{f\in\{q,l,u,d,e\}}i\bar f\gamma^\mu D_\mu f -\bigl[(\bar u y_u^\dagger q_\beta \epsilon^{\beta\alpha} + \bar q^\alpha \VCKM y_d d + \bar l^\alpha y_e e)H_\alpha+\text{h.c.}\bigr],
\eeqa{LSM}
where $q=(u_L,d_L)$, $l=(\nu,e_L)$, $u=u_R$, $d=d_R$, $e=e_R$. All the gauge-eigenstate fermion fields are also mass eigenstates except $d_L=\VCKM d'_L$ where $d'_L$ is a mass eigenstate. In the last term, $\alpha$ and $\beta$ are $SU(2)_L$ indices of the doublet fields, while generation indices are implicitly summed over. The Yukawa matrices $y_u$, $y_d$, $y_e$ are real and diagonal in generation space, and should not be confused with the hypercharges
\beq
\{Y_q, Y_l, Y_u, Y_d, Y_e\} = \{\frac{1}{6}, -\frac{1}{2}, \frac{2}{3}, -\frac{1}{3}, -1\}.
\eeqn
The sign conventions are, for example,
\bseq
\beqa
G^A_{\mu\nu} &=& \partial_\mu G^A_\nu -\partial_\nu G^A_\mu +g_sf^{ABC}G^B_\mu G^C_\nu, \\
D_\mu q &=& (\partial_\mu-ig_sT^AG^A_\mu-ig\frac{\sigma^a}{2}W^a_\mu-ig'Y_qB_\mu) q, \\
W^3_\mu &=& \cw Z_\mu +\sw A_\mu,\quad B_\mu = -\sw Z_\mu +\cw A_\mu, \\
\cw &=& \frac{g}{\sqrt{g^2+g'^2}} = \frac{e}{g'},\quad \sw = \frac{g'}{\sqrt{g^2+g'^2}}= \frac{e}{g}.
\eeqan
\eseqn

In the SMEFT, \eqref{LSM} is supplemented by the complete set of dimension-6 operators. We work with the Warsaw basis~\cite{Grzadkowski:2010es}, and adopt the conventions in~\cite{Grzadkowski:2010es} for the effective operators. In this basis, there are 9 CP-even bosonic operators,
\beqa
&& Q_{HW}=|H|^2W^a_{\mu\nu}W^{a\mu\nu},\quad Q_{HB}=|H|^2B_{\mu\nu}B^{\mu\nu},\quad Q_{HG}=|H|^2G^A_{\mu\nu}G^{A\mu\nu} \CR
&& Q_{HWB}=H^\dagger\sigma^a HW^a_{\mu\nu}B^{\mu\nu},\quad Q_W=\epsilon^{abc}W_\mu^{a\nu}W_\nu^{b\rho}W_\rho^{c\mu},\quad Q_G=f^{ABC}G_\mu^{A\nu}G_\nu^{B\rho}G_\rho^{C\mu}, \CR
&& Q_{HD}=|H^\dagger D_\mu H|^2,\quad Q_{H\square}=|H|^2\square|H|^2,\quad Q_H=|H|^6.
\eeqan
The LO universal theories EFT Lagrangian \eqref{Lu} consists of these 9 operators and 7 linear combinations of fermionic operators listed in table~\ref{tab:Qf}: $Q_{HJW}$ and $Q_{HJB}$ from 7 of the 8 $\psi^2H^2D$-class operators
\beqa
&& Q_{Hq}^{(3)} = (iH^\dagger\sigma^a\Dlr_\mu H)(\bar q\gamma^\mu\sigma^a q),\quad Q_{Hl}^{(3)} = (iH^\dagger\sigma^a\Dlr_\mu H)(\bar l\gamma^\mu\sigma^a l), \CR
&& Q_{Hq}^{(1)} = (iH^\dagger\Dlr_\mu H)(\bar q\gamma^\mu q),\quad Q_{Hu} = (iH^\dagger\Dlr_\mu H)(\bar u\gamma^\mu u),\quad Q_{Hd} = (iH^\dagger\Dlr_\mu H)(\bar d\gamma^\mu d), \CR
&& Q_{Hl}^{(1)} = (iH^\dagger\Dlr_\mu H)(\bar l\gamma^\mu l),\quad Q_{He} = (iH^\dagger\Dlr_\mu H)(\bar e\gamma^\mu e);
\eeqa{psi2H2D}
$Q_y$ from the 3 $\psi^2H^3$-class operators
\beq
Q_{uH} = |H|^2 \bar q_\beta u \epsilon^{\beta\alpha} H^\dagger_\alpha,\quad Q_{dH} = |H|^2 \bar q^\alpha d H_\alpha,\quad Q_{eH} = |H|^2 \bar l^\alpha e H_\alpha;
\eeq{psi2H3}
and $Q_{2JW}$, $Q_{2JB}$, $Q_{2JB}$, $Q_{2y}$ from 23 of the 25 four-fermion operators
\beqa
&& Q_{qq}^{(3)} = (\bar q\gamma_\mu\sigma^a q)(\bar q\gamma^\mu\sigma^a q),\quad Q_{lq}^{(3)} = (\bar l\gamma_\mu\sigma^a l)(\bar q\gamma^\mu\sigma^a q), \CR
&& Q_{qq}^{(1)} = (\bar q\gamma_\mu q)(\bar q\gamma^\mu q),\quad Q_{lq}^{(1)} = (\bar l\gamma_\mu l)(\bar q\gamma^\mu q),\quad Q_{ll} = (\bar l\gamma_\mu l)(\bar l\gamma^\mu l),\CR
&& Q_{uu} = (\bar u\gamma_\mu u)(\bar u\gamma^\mu u),\quad Q_{dd} = (\bar d\gamma_\mu d)(\bar d\gamma^\mu d),\quad Q_{ee} = (\bar e\gamma_\mu e)(\bar e\gamma^\mu e), \CR
&& Q_{qu}^{(1)} = (\bar q\gamma_\mu q)(\bar u\gamma^\mu u),\quad Q_{qd}^{(1)} = (\bar q\gamma_\mu q)(\bar d\gamma^\mu d),\quad Q_{qe} = (\bar q\gamma_\mu q)(\bar e\gamma^\mu e), \CR
&& Q_{lu} = (\bar l\gamma_\mu l)(\bar u\gamma^\mu u),\quad Q_{ld} = (\bar l\gamma_\mu l)(\bar d\gamma^\mu d),\quad Q_{le} = (\bar l\gamma_\mu l)(\bar e\gamma^\mu e), \CR
&& Q_{ud}^{(1)} = (\bar u\gamma_\mu u)(\bar d\gamma^\mu d),\quad Q_{eu} = (\bar e\gamma_\mu e)(\bar u\gamma^\mu u),\quad Q_{ed} = (\bar e\gamma_\mu e)(\bar d\gamma^\mu d), \CR
&& Q_{qu}^{(8)} = (\bar q\gamma_\mu T^A q)(\bar u\gamma^\mu T^A u),\quad Q_{qd}^{(8)} = (\bar q\gamma_\mu T^A q)(\bar d\gamma^\mu T^A d),\quad Q_{ud}^{(8)} = (\bar u\gamma_\mu T^A u)(\bar d\gamma^\mu T^A d), \CR
&& Q_{quqd}^{(1)} = (\bar q_\alpha u)\epsilon^{\alpha\beta}(\bar q_\beta d),\quad Q_{lequ}^{(1)} = (\bar l_\alpha e)\epsilon^{\alpha\beta}(\bar q_\beta u),\quad Q_{ledq} = (\bar l_\alpha e)(\bar d q^\alpha).
\eeqa{psi4}
The $9+8+3+25=45$ operators mentioned above (42 explicitly listed), plus the 8 $\psi^2XH$-class operators, constitute the 53 independent CP-even, baryon-number-conserving dimension-6 operators (ref.~\cite{Grzadkowski:2010es} further lists 6 CP-odd operators, making the total number 59). Generation indices have been suppressed in \eqref{psi2H2D}, \eqref{psi2H3}, \eqref{psi4}, for which our conventions are, e.g.\
\bseq
\beqa
&& [C_{Hl}^{(1)}]_{ij}[Q_{Hl}^{(1)}]_{ij} = [C_{Hl}^{(1)}]_{ij}(iH^\dagger\Dlr_\mu H)(\bar l_i\gamma^\mu l_j), \\ 
&& [C_{lq}^{(1)}]_{ijkl}[Q_{lq}^{(1)}]_{ijkl} = [C_{lq}^{(1)}]_{ijkl}(\bar l_i\gamma_\mu l_j)(\bar q_k\gamma^\mu q_l).
%[C_{ll}]_{ijkl}[Q_{ll}]_{ijkl} = [C_{ll}]_{ijkl}(\bar l_i\gamma_\mu l_j)(\bar l_k\gamma^\mu l_l).
\eeqan
\eseqn

When using the results in~\cite{Jenkins:2013zja,Jenkins:2013wua,Alonso:2013hga}, we need to flip the signs of the gauge couplings $g_s$, $g$, $g'$, and replace the Yukawa matrices $Y_u^\dagger$, $Y_d^\dagger$, $Y_e^\dagger$ in these references by $y_u$, $\VCKM y_d$, $y_e$, respectively to conform with our notation.

%%%
\section{Universal parameters from the effective Lagrangian}
\label{app:up}

The 16 independent universal parameters listed in table~\ref{tab:up} can be identified with coefficients of terms in the effective Lagrangian, when the latter is written in the electroweak symmetry broken phase in the unitary gauge, i.e.\ $H=\frac{1}{\sqrt{2}}(0,v+h)$, and the SM fields and parameters are redefined to satisfy the oblique parameters defining conditions~\cite{Barbieri:2004qk,Wells:2015uba}. Denoting these properly-redefined fields and parameters with bars, we have
\beqa
\L_{\text{universal}} &=& \Bigl(\frac{\bar g\bar v}{2}\Bigr)^2 \bar W^+_\mu \bar W^{-\mu} +(1-\hat T)\frac{1}{2} \Bigl(\frac{\bar g\bar v}{2\cwb}\Bigr)^2 \bar Z_\mu \bar Z^\mu \CR
&& -\frac{1}{2}\bar G^A_\mu \hat K^{\mu\nu} \bar G^A_\nu -\bar W^+_\mu \hat K^{\mu\nu} \bar W^-_\nu -\frac{1}{2} \bar W^3_\mu \hat K^{\mu\nu} \bar W^3_\nu -\hat S\frac{\swb}{\cwb} \bar W^3_\mu \hat K^{\mu\nu} \bar B_\nu -\frac{1}{2} \bar B_\mu \hat K^{\mu\nu} \bar B_\nu \CR
&& -\frac{1}{m_W^2}\Bigl[Z \frac{1}{2}\bar G^A_\mu \hat K^{2\mu\nu} \bar G^A_\nu +W \Bigl(\bar W^+_\mu \hat K^{2\mu\nu} \bar W^-_\nu +\frac{1}{2}\bar W^3_\mu \hat K^{2\mu\nu} \bar W^3_\nu\Bigr) +Y \frac{1}{2}\bar B_\mu \hat K^{2\mu\nu} \bar B_\nu \Bigr] \CR
&& +i\bar g \Bigl\{ (\bar W^+_{\mu\nu}\bar W^{-\mu}-\bar W^-_{\mu\nu}\bar W^{+\mu}) \bigl[ (1+\Dgzb) \cwb \bar Z^\nu +\swb\bar A^\nu \bigr] \CR
&&\qquad +\frac{1}{2}\bar W^+_{[\mu,}\bar W^-_{\nu]} \bigl[ (1+\Dkapzb) \cwb\bar Z^{\mu\nu} +(1 +\Dkapab) \swb\bar A^{\mu\nu} \bigr] \CR
&&\qquad +\frac{\lamab}{m_W^2} \bar W^{+\nu}_\mu \bar W^{-\rho}_\nu (\cwb\bar Z_\rho ^{\,\,\,\mu}+\swb\bar A_\rho^{\,\,\,\mu}) \Bigr\} +\frac{W}{m_W^2}\hat K\circ \L_{\bar W\bar W\bar V}^{\SM} \CR
&&\qquad +\L_{\bar G^3}^{\SM} -\frac{\lamgb}{m_W^2}\frac{\bar g_s}{6}f^{ABC}\bar G_\mu^{A\nu}\bar G_\nu^{B\rho}\bar G_\rho^{C\mu} +\frac{Z}{m_W^2}\hat K\circ \L_{\bar G^3}^{\SM} \CR
% h
&& +\frac{1}{2}\partial_\mu \bar h \partial^\mu \bar h -\frac{1}{2}(2\bar\lambda\bar v^2) \bar h^2 -(1+\Delta\kappa_3) \bar\lambda\bar v\bar h^3 \CR
% hf
&& -\Bigl[1 +(1+\DkapFb)\frac{\bar h}{\bar v} +\Bigl(\frac{3}{2}\DkapFb-\frac{1}{2}\DkapVb\Bigr)\frac{\bar h^2}{\bar v^2}\Bigr] \sum_{f'} \frac{\bar y_{f'} \bar v}{\sqrt{2}} \bar f' f' \CR
% hV
&& +(1+\DkapVb)\frac{2\bar h}{\bar v} \biggl[\Bigl(\frac{\bar g\bar v}{2}\Bigr)^2\bar W^+_\mu\bar W^{-\mu} +(1-2\hat T)\frac{1}{2}\Bigl(\frac{\bar g\bar v}{2\cwb}\Bigr)^2\bar Z_\mu\bar Z^\mu\biggr]\CR
&&\qquad +(1+4\DkapVb)\frac{\bar h^2}{\bar v^2} \biggl[\Bigl(\frac{\bar g\bar v}{2}\Bigr)^2\bar W^+_\mu\bar W^{-\mu} +(1-6\hat T)\frac{1}{2}\Bigl(\frac{\bar g\bar v}{2\cwb}\Bigr)^2\bar Z_\mu\bar Z^\mu\biggr] \CR
&& +\Bigl(\frac{\bar h}{\bar v}+\frac{\bar h^2}{2\bar v^2}\Bigr) \Bigl[ f_{gg}\frac{\bar g_s^2}{4} \partial_{[\mu,}\bar G^A_{\nu]} \partial^{[\mu,}\bar G^{A\nu]} +f_{ww}\frac{\bar g^2}{2} \bar W^+_{\mu\nu}\bar W^{-\mu\nu} +f_{zz}\frac{\bar g^2}{4\cwb^2}\bar Z_{\mu\nu}\bar Z^{\mu\nu} \CR
&&\qquad +f_{z\gamma}\frac{\bar g\bar g'}{2}\bar Z_{\mu\nu}\bar A^{\mu\nu} +f_{\gamma\gamma}\frac{\bar e^2}{4} \bar A_{\mu\nu}\bar A^{\mu\nu} +f_{w\square}\bar g^2(\bar W^-_\mu\partial_\nu \bar W^{+\mu\nu}+\text{h.c.}) \CR
&&\qquad +f_{z\square}\bar g^2\bar Z_\mu\partial_\nu\bar Z^{\mu\nu} +f_{\gamma\square}\bar g\bar g'\bar Z_\mu\partial_\nu\bar A^{\mu\nu}\Bigr] \CR
&& +c_{2y} J_{y\alpha}^\dagger J_y^\alpha +\sum_{f} i\bar f\gamma^\mu D_\mu f +\O(\bar V^4, \bar h^4, \bar h^3 f^2, \bar h^3\bar V^2, \bar h\bar V^3).
\eeqa{Lu-up}
With the standard notation, $W^\pm_\mu=\frac{1}{\sqrt{2}}(W^1_\mu \mp iW^2_\mu)$, $W^\pm_{\mu\nu} = \partial_{[\mu,}W^\pm_{\nu]}$, $Z_{\mu\nu} = \partial_{[\mu,}Z_{\nu]}$, $A_{\mu\nu} = \partial_{[\mu,}A_{\nu]}$, where $(\dots)_{[\mu,\nu]}\equiv(\dots)_{\mu\nu}-(\dots)_{\nu\mu}$ denotes an antisymmetric tensor. In \eqref{Lu-up} we have defined
\beq
\hat K^{\mu\nu} \equiv -g^{\mu\nu}\partial^2+\partial^\mu\partial^\nu, \quad \hat K^{2\mu\nu} \equiv \hat K^{\mu\rho}\hat K_\rho^{\,\,\,\nu}.
\eeqn
The action of $\hat K\circ$ follows the product rule, e.g.\
\beqa
&&\hat K\circ (W^+_{\mu\nu}W^{-\mu}Z^\nu) = \hat K\circ (\partial_{[\mu,}W^+_{\nu]}W^{-\mu}Z^\nu) \CR
&&\qquad =\partial_{[\mu,}(\hat K W^+)_{\nu]} W^{-\mu}Z^\nu +\partial_{[\mu,}W^+_{\nu]}(\hat K W^-)^\mu Z^\nu +\partial_{[\mu,}W^+_{\nu]}W^{-\mu}(\hat K Z)^\nu,
\eeqan
where $(\hat K W^+)_{\nu}=\hat K_{\nu\rho} W^{+\rho}$, etc. Also, we have used $f'$ to denote mass-eigenstate fields and $f$ to denote gauge-eigenstate fields for the SM fermions. The reader is referred to~\cite{Wells:2015uba} for details of the reduction from \eqref{Lu} to \eqref{Lu-up}.

Other parameters appearing in \eqref{Lu-up} depend on the universal parameters as follows,
\bseq
\beqa
\Dkapzb &=& \Dgzb -\frac{\sw^2}{\cw^2}\Dkapab, \\
f_{ww} &=& f_{z\gamma} +\sw^2f_{\gamma\gamma} +\frac{2}{g^2}\Dkapab, \\
f_{zz} &=& (\cw^2-\sw^2)f_{z\gamma} +\cw^2\sw^2f_{\gamma\gamma} +\frac{2}{g^2}\Dkapab, \\
f_{w\square} &=& -\frac{2\cw^2}{g^2}\Dgzb, \\
f_{z\square} &=& -\frac{2}{g^2}\Bigl[(\cw^2-\sw^2)\Dgzb +\frac{\sw^2}{\cw^2}(\Dkapab-\hat S)\Bigr], \\
f_{\gamma\square} &=& -\frac{2}{g^2}(2\cw^2\Dgzb-\Dkapab+\hat S).
\eeqan
\eseqn

%%%
\section{Higgs basis couplings in the Warsaw basis}
\label{app:hb}

In this appendix we collect definitions of the Higgs basis couplings~\cite{HiggsBasis} that are relevant for our discussion of RG effects in sections~\ref{sec:ew} and~\ref{sec:yukawa}, and their expressions in terms of the Warsaw basis Wilson coefficients. Note that a slightly different notation is adopted in~\cite{HiggsBasis} compared with the original Warsaw basis paper~\cite{Grzadkowski:2010es}. In particular, $C_{HWB}$, $C_{HD}$, $C_{H\square}$ in~\cite{Grzadkowski:2010es} (and also in the present paper) translate into $gg'c_{WB}$, $-4c_T$, $-c_H-c_T$, respectively, in~\cite{HiggsBasis}. Also, the $\psi^2H^3$-class operators %$Q_{uH}$, $Q_{dH}$, $Q_{eH}$ 
are defined differently in~\cite{HiggsBasis} than in~\cite{Grzadkowski:2010es}.

With the SM fields and parameters properly redefined to satisfy the Higgs basis defining conditions~\cite{HiggsBasis,Wells:2015uba}, the SMEFT contains the following terms for the charged-current (CC) and neutral-current(NC) interactions of the SM fermions,
\bseq
\beqa
\hspace{-2ex}\L_{\text{CC}} &=& \frac{\hat g}{\sqrt{2}}\Bigl\{\hat W^+_\mu\Bigl[\bigl(\delta_{ij}+[\dgL^{Wq}]_{ij}\bigr)\bar u_{L,i}\gamma^\mu d_{L,j} +\bigl(\delta_{ij}+[\dgL^{Wl}]_{ij}\bigr)\bar \nu_i\gamma^\mu e_{L,j}\Bigr] +\text{h.c.}\Bigr\}, \\
\hspace{-2ex}\L_{\text{NC}} &=& \sum_f \Bigl[\frac{\hat e}{\cwh\swh}\hat Z_\mu\bigl((T^3_f-Q_f\swh^2)\delta_{ij} +[\delta g_{L/R}^{Zf}]_{ij}\bigr) +\hat e\hat A_\mu Q_f\delta_{ij} \Bigr] \bar f_i\gamma^\mu f_j,
\eeqan
\eseq{CCNC}
where $\dgL^{Zf}$ and $\dgR^{Zf}$ apply to $f\in\{u_L,d_L,e_L,\nu\}$ and $f\in\{u_R,d_R,e_R\}$, respectively, and $T^3_f=0$ is assumed for $f\in\{u_R,d_R,e_R\}$. The fields and parameters satisfying the Higgs basis defining conditions have been denoted with hats. They are in general different from the barred fields in \eqref{Lu-up} which satisfy the oblique parameters defining conditions; see~\cite{Wells:2015uba}.

We have followed the conventions in~\cite{HiggsBasis} for the definitions of the Higgs basis couplings in \eqref{CCNC}, with the exception that our $\dgL^{Wq}$ is defined with respect to the gauge-eigenstate fields rather than the mass-eigenstate fields. In our notation, $[\dgL^{Wl}]_{ij}$ and $[\dgL^{Wq}]_{ij}$ are given by \eqref{dgLWfmu}, while the anomalous $Zf\bar f$ couplings are 
\bseq
\beqa
%[\dgL^{Wl}(\mu)]_{ij} &\equiv& [C_{Hl}^{(3)}(\mu)]_{ij} -\frac{\cw\sw}{\cw^2-\sw^2} C_{HWB}(\mu) -\frac{\cw^2}{\cw^2-\sw^2}C_0(\mu),\\%\label{dgLWlmu} \\
%{[}\dgL^{Wq}(\mu)]_{ij} &\equiv& [C_{Hq}^{(3)}(\mu)]_{ij} -\frac{\cw\sw}{\cw^2-\sw^2} C_{HWB}(\mu) -\frac{\cw^2}{\cw^2-\sw^2}C_0(\mu),\\%\label{dgLWqmu} \\
[\dgL^{Zu}]_{ij} &\equiv& \frac{1}{2}[C_{Hq}^{(3)}]_{ij} -\frac{1}{2}[C_{Hq}^{(1)}]_{ij} -\frac{2\cw\sw}{3(\cw^2-\sw^2)} C_{HWB} -\frac{3\cw^2+\sw^2}{6(\cw^2-\sw^2)}C_0, \\
{[}\dgR^{Zu}]_{ij} &\equiv& -\frac{1}{2}[C_{Hu}]_{ij} -\frac{2\cw\sw}{3(\cw^2-\sw^2)} C_{HWB} -\frac{2\sw^2}{3(\cw^2-\sw^2)}C_0, \\
{[}\dgL^{Zd}]_{ij} &\equiv& -\frac{1}{2}[C_{Hq}^{(3)}]_{ij} -\frac{1}{2}[C_{Hq}^{(1)}]_{ij} +\frac{\cw\sw}{3(\cw^2-\sw^2)} C_{HWB} +\frac{3\cw^2-\sw^2}{6(\cw^2-\sw^2)}C_0, \\
{[}\dgR^{Zd}]_{ij} &\equiv& -\frac{1}{2}[C_{Hd}]_{ij} +\frac{\cw\sw}{3(\cw^2-\sw^2)} C_{HWB} +\frac{\sw^2}{3(\cw^2-\sw^2)}C_0, \\
{[}\dgL^{Ze}]_{ij} &\equiv& -\frac{1}{2}[C_{Hl}^{(3)}]_{ij} -\frac{1}{2}[C_{Hl}^{(1)}]_{ij} +\frac{\cw\sw}{\cw^2-\sw^2} C_{HWB} +\frac{1}{2(\cw^2-\sw^2)}C_0, \\
{[}\dgR^{Ze}]_{ij} &\equiv& -\frac{1}{2}[C_{He}]_{ij} +\frac{\cw\sw}{\cw^2-\sw^2} C_{HWB} +\frac{\sw^2}{\cw^2-\sw^2}C_0, \\
{[}\dgL^{Z\nu}]_{ij} &\equiv& \frac{1}{2}[C_{Hl}^{(3)}]_{ij} -\frac{1}{2}[C_{Hl}^{(1)}]_{ij} -\frac{1}{2}C_0,
\eeqan
\eseq{dgZf}
where $C_0$ is the Wilson coefficient combination defined in \eqref{C0def}, and is identified with $\delta v-c_T$ in the notation of~\cite{HiggsBasis}.
%\beq
%C_{0}(\mu) \equiv \frac{1}{4} \Bigl\{C_{HD}(\mu) +2 \bigl([C_{Hl}^{(3)}(\mu)]_{11} +[C_{Hl}^{(3)}(\mu)]_{22}\bigr) -\bigl([C_{ll}(\mu)]_{1221}+[C_{ll}(\mu)]_{2112}\bigr)\Bigr\}
%\eeqn

The Higgs boson couplings to SM fermions, on the other hand, are given by
\beq
\L_{hff} = -\frac{\hat h}{v}\sum_{f'=u,d,e}\sum_{i,j}\sqrt{m_{f'_i} m_{f'_j}} \bar f'_i \Bigl(\delta_{ij}+[\delta y_{f'}]_{ij}(\cos\phi_{ij}^{f'}-i\sin\phi_{ij}^{f'}\gamma^5)\Bigr) f'_j,
\eeqn
For general flavor structures of the $\psi^2H^3$-class operators, the fermion mass matrices need to be rediagonalized to define the mass eigenstates $f'_i$. In universal theories (with RG effects included), and in the approximation \eqref{VCKMapprox}, the third-generation fermions are not affected by this rotation. We have $\phi_{ij}^f=0$, and
\bseq
\beqa
\delta y_t &=& [\delta y_u]_{33} = -\frac{[C_{uH}]_{33}}{y_t} +C_{H\square} -C_0, \\
\delta y_b &=& [\delta y_d]_{33} = -\frac{[C_{dH}]_{33}}{y_b} +C_{H\square} -C_0, \\
\delta y_\tau &=& [\delta y_e]_{33} = -\frac{[C_{eH}]_{33}}{y_\tau} +C_{H\square} -C_0.
\eeqan
\eseq{dyf}


\end{document}